\documentclass[twocolumn,notitlepage,aps,superscriptaddress]{revtex4-1}
\usepackage{tikz}
\usepackage{cmap}
\usepackage{color}
\usepackage[english]{babel}
\usepackage[utf8]{inputenc}
\usepackage{hyperref}
\usepackage{graphicx}
\usepackage{amsmath}
\usepackage{amssymb}
\usepackage{amsfonts}
\usepackage{amsthm}
\usepackage{textcomp}
\usepackage{scrpage2}
\usepackage[braket]{qcircuit}
\usepackage{verbatim}
\usepackage{mathtools}
\begin{document}
\title{Lowering qubit requirements for quantum simulations of fermionic systems}
\begin{abstract}
The mapping of  fermionic states onto qubit states, as well as the mapping of fermionic Hamiltonian into quantum gates enables us to simulate electronic systems with a quantum computer. Benefiting the understanding of many-body systems in chemistry and physics, quantum simulation  is one of the great promises of the coming age of quantum computers.
One challenge in realizing simulations on near-term quantum devices is the large number of qubits required by such mappings.
In this work, we develop methods that allow us to trade-off qubit requirements against the complexity of the resulting quantum circuit. We first show that any classical code used to map the state of a fermionic Fock space to qubits gives rise to a mapping of fermionic models to quantum gates. As an illustrative example, we present a mapping based on a non-linear classical error correcting code, which leads to significant qubit savings albeit at the expense of additional quantum gates.
We proceed to use this framework to present a number of simpler mappings that lead to qubit savings with only a very modest increase in gate difficulty.
We discuss the role of symmetries such as particle conservation, and savings that could be obtained if an experimental platform could easily realize multi-controlled gates.
\end{abstract}

\author{Mark Steudtner}\affiliation{Instituut-Lorentz, Universiteit Leiden, P.O. Box 9506, 2300 RA Leiden, The Netherlands}
\affiliation{QuTech, Delft University of Technology, Lorentzweg 1, 2628 CJ Delft, The Netherlands}
\author{Stephanie Wehner}
\affiliation{QuTech, Delft University of Technology, Lorentzweg 1, 2628 CJ Delft, The Netherlands}
\date{\today} \maketitle

\newcommand{\bbs}[1]{\boldsymbol{#1}}
\newcommand{\moto}{\;\mathrm{mod}\;2}
\newcommand{\party}{checksum code }
\newcommand{\Party}{Checksum code }
\newcommand{\parties}{checksum codes }
\newcommand{\Parties}{Checksum codes }
\newcommand{\hamming}[1]{\mathrm{w_H}\left( #1 \right)}
\newcommand{\enc}{\mathbf{e}}
\newcommand{\dec}{\mathbf{d}}
\newcommand{\Complex}{\mathbb{C}}
\newcommand{\id}{\mathbb{I}}
\newcommand{\steph}[1]{{\textcolor{red}{Steph: #1}}}
\section{Introduction}
Simulating quantum systems on a quantum computer is one of the most promising applications of small scale quantum computers~\cite{olson2017quantum}.
Significant efforts have gone into the theoretical development of simulation algorithms~\cite{bravyi2002fermionic,aspuru2005simulated,whitfield2011simulation,hastings2014improving,babbush2014adiabatic}, and their experimental demonstrations~\cite{lanyon2010towards,du2010nmr,peruzzo2014variational,wang2015quantum,o2016scalable,kandala2017hardware}. Resource estimates~\cite{wecker2013can,poulin2014trotter,babbush2015chemical}, such as for example for FeMoCo, a model for the nitrogenase enzyme, indicate that simulations of relevant chemical systems may be achieved with relatively modest quantum computing resources~\cite{reiher2017elucidating} in comparison to many standard quantum algorithms~\cite{shor1999polynomial,harrow2009quantum}.

One essential component in realizing simulations of fermionic models on quantum computers is the representation of such models
in terms of qubits and quantum gates.
Following initial simulation schemes for fermions hopping on a lattice~\cite{abrams1997simulation}, more recent proposals used the  Jordan-Wigner~\cite{wigner1928uber}
transform~\cite{somma2002simulating,somma2003quantum,aspuru2005simulated,lanyon2010towards}, the Verstraete-Cirac mapping~\cite{verstraete2005mapping}, or the Bravyi-Kitaev transform~\cite{bravyi2002fermionic} to find a suitable representation.
Specifically, the task of all such representations is two-fold. First, we seek a mapping from states in the
fermionic Fock space of $N$ sites to the space of $n$ qubits. The fermionic Fock space is spanned by $2^N$ basis vectors
$\ket{\nu_1,\ldots,\nu_N}$ where $\nu_j\in\{0,1\}$ indicates the presence ($\nu_j = 1)$ or absence ($\nu_j = 0$) of a spinless fermionic particle at orbital $j$ \footnote{We slightly abuse the nomenclature of quantum chemistry and  molecular physics in merging spatial and spin quantum numbers into one index $j$, and use it as a label for  what we call now the $j$-th orbital.}.
Such a mapping $\enc: \mathbb{Z}_2^{\otimes N} \to \mathbb{Z}_2^{\otimes n}$ is also called an \emph{encoding}~\cite{seeley2012bravyi}.
An example of such an encoding is the trivial one in which $n=N$ and qubits are used to represent the binary string $\bbs{\nu}=\left(\nu_1, \, ...\,, \, \nu_N\right)^\top$. That is,
\begin{align}\label{eq:trivial}
\ket{\bbs{\omega}} = \ket{\enc\left(\bbs{\nu}\right)} = \bigotimes_{j=1}^n \ket{\omega_j}\,
\end{align}
where $\omega_j = \nu_j$ in the standard basis $\{\ket{0},\ket{1}\}$.

Second, we need a way to simulate the dynamics of fermions on these $N$ orbitals. These dynamics can be modeled entirely in  terms of the annihilation and creation
operators $c_j$ and $c_j^\dagger$ that act on the fermionic Fock space as
\begin{align} \label{eq:stateaction1}
c_{i_m}^\dagger \; c^\dagger_{i_1}...\; c^\dagger_{i_{m-1}} c_{i_m}^\dagger c^\dagger_{i_{m+1}} ...\; c^\dagger_{i_{M}} \ket{\Theta} = 0 \\
c_{i_m}^{\;} \; c^\dagger_{i_1}...\; c^\dagger_{i_{m-1}}c^\dagger_{i_{m+1}} ...\; c^\dagger_{i_{M}} \ket{\Theta} = 0 \end{align}\begin{align}
c_{i_m}^{\;} \;  c^\dagger_{i_1}...\; c^\dagger_{i_{m-1}} c^\dagger_{i_{m}}    c^\dagger_{i_{m+1}} ...\; c^\dagger_{i_{M}} \ket{\Theta}   \nonumber \\ = (-1)^{m-1} \; c^\dagger_{i_1} ...\;c^\dagger_{i_{m-1}}  c^\dagger_{i_{m+1}} ...\; c^\dagger_{i_{M}} \ket{\Theta} \end{align}\begin{align}
c_{i_m}^\dagger \; c^\dagger_{i_1}...\; c^\dagger_{i_{m-1}}  c^\dagger_{i_{m+1}} ...\; c^\dagger_{i_{M}} \ket{\Theta}   \nonumber \\  = (-1)^{m-1}\; c^\dagger_{i_1}... \;c^\dagger_{i_{m-1}} c^\dagger_{i_{m}} c^\dagger_{i_{m+1}} ... \;c^\dagger_{i_{M}} \ket{\Theta}  \label{eq:stateaction4} \,
\end{align}

where $\ket{\Theta}$ is the fermionic vacuum and $\left\lbrace  i_1 ,\, ...\, ,\, i_M \right\rbrace\subseteq \lbrace 1,\, ...\, ,\, N \rbrace$.
The operators satisfy the anti-commutation relations
\begin{align}
\label{eq:antiraw}
\left[ c^{\vphantom{\dagger}}_i,c^{\vphantom{\dagger}}_j\right]_+ = 0 , \quad \left[ c^\dagger_i,c^\dagger_j\right]_+=0, \quad \left[ c_i,c^\dagger_j\right]_+ = \delta_{ij}\ ,
\end{align}
with $[A,B]_+ = AB + BA$. Mappings of the operators $c_j$ to qubits typically use the Pauli matrices $X$, $Z$, and $Y$ acting on one qubit, characterized
by their anti-commutation relations $[P_i,P_j]_+ = 2 \delta_{ij} \id$ for all $P_i\in\mathcal{P}=\{X,Z,Y\}$.
An example of such a mapping is the Jordan-Wigner transform~\cite{wigner1928uber} given by
\begin{align}
c_j \;&\hat{=}\; Z^{\otimes j-1} \otimes \sigma^- \otimes \id^{\otimes n-j} \\
c_j^{\dagger} \;&\hat{=}\; Z^{\otimes j-1} \otimes \sigma^+ \otimes \id^{\otimes n-j}
\end{align}
where
\begin{align}
\sigma^- = \ket{0}\!\bra{1} = \frac{1}{2}\left(X + i Y\right)\ ,\\
\sigma^+ = \ket{1}\!\bra{0} = \frac{1}{2}\left(X - i Y\right)\ .
\end{align}
It is easily verified that together with the trivial encoding~\eqref{eq:trivial} this transformation satisfies
the desired properties~\eqref{eq:stateaction1}-\eqref{eq:antiraw} and can hence be used to represent fermionic models with qubit systems.

In order to assess the suitability of an encoding scheme for the simulation of fermionic models on a quantum computer, a number of parameters
are of interest. The first is the total number of qubits $n$ needed in the simulation. Second, we may care about the gate size of the operators $c_j$ and
$c_j^\dagger$ when mapped to qubits. In its simplest form, this problem concerns the total number of qubits on which these operators do not act trivially, that is,
the number of qubits $L$, on which an operator $P_j \in \mathcal{P}$ acts instead of the identity $\id$, sometimes called the Pauli length.
Different transformations can lead to dramatically different performance
with respect to these parameters. For both the Jordan-Wigner as well as the Bravyi-Kitaev transform $n=N$, but we have $L=O(n)$ for the first, while
$L=O(\log n)$ for the second.
We remark that in experimental implementations we typically do not only care about the absolute number $L$, but
rather the specific gate size and individual
difficulty of the qubit gates each of which may be easier or harder to realize in a specific experimental architecture.
Finally, we remark that instead of looking for a mapping for individual operators $c_j$ we may instead opt to map pairs (or higher order terms) of such operators at once, or even look to represent sums of such operators.

\subsection{Results}
Here, we propose a general family of mappings of fermionic models to qubit systems
and quantum gates that allow us to trade off the necessary number of qubits
$n$ against the difficulty of implementation as parametrized by $L$, or more complicated quantum gates such as $\textsc{CPhase}$.
Ideally, one would of course like both the number of qubits, as well as the gate size to be small.
We show that our mappings can lead to significant savings in qubits for a variety of examples (see Table~\ref{tab:othertable}) as compared to the Jordan-Wigner transform for instance, at the expense of greater complexity in realizing the required gates.
The latter may lead to an increased time required for the simulation depending on which gates are easy to realize in a particular quantum computing architecture.

At the heart of our efforts is an entirely general construction of the creation and annihilation operators in~\eqref{eq:stateaction1} given an arbitrary
 encoding $\enc$ and the corresponding decoding $\dec$. As one might expect, this construction is not efficient for every choice of encoding $\enc$ or decoding $\dec$. However, for linear encodings $\enc$, but possibly non-linear decodings $\dec$, they can take on a very nice form.  While in principle any classical
code with the same properties can be shown to yield such mappings, we provide an appealing example of how a classical code of fixed Hamming
weight~\cite{tian2007constant} can be used to give an interesting mapping.

Two other approaches allow us to  be more modest with the algorithmic depth in either accepting a qubit saving that is linear with $N$, or just saving a fixed amount of qubits for hardly any cost at all.  \\
In previous works, trading quantum resources  has been addressed for general algorithms \cite{bravyi2016trading}, and quantum simulations \cite{moll2016optimizing, bravyi2017tapering,romero2016quantum}. In the two works of Moll et al. and Bravyi et al.,  qubit requirements are reduced  with a scheme that is different from ours. A qubit Hamiltonian is first obtained with e.g. the Jordan-Wigner transform, then unitary operations are applied to it in order taper qubits off successively. The notion that our work is based on,  was first introduced in \cite{bravyi2017tapering} by Bravyi et al., for linear en- and decodings.
With the generalization of this method, we hope to make the goal of qubit reduction more attainable in reducing the effort to do so. The reduction method is mediated by nonlinear codes, of  which we  provide different types to choose from.  The transform of the Hamiltonian is straight-forward from there on, and we give explicit recipes for arbitrary codes.
We can summarize our contributions as follows.
\begin{itemize}
\item We show that for any encoding $\enc: \mathbb{Z}_2^{\otimes N} \to \mathbb{Z}_2^{\otimes n}$ there exists a mapping of Fermionic models to quantum gates. For the special case that this encoding is linear, our procedure can be understood as a slightly modified version of the perspective taken in~\cite{seeley2012bravyi}.
This gives a systematic way to employ classical codes for obtaining such mappings.
\item Using particle-conservation symmetry, we develop 3 types of codes that save a constant, linear and exponential amount of qubits (see Table \ref{tab:othertable} and Sections \ref{subsubsec:party}-\ref{subsubsec:mentcodes}). An example from classical coding theory~\cite{tian2007constant} is used to obtain significant qubit savings (here called the binary addressing code),
at the expense of increased gate difficulty (unless the architecture would easily support multi-controlled gates).
\item The codes developed are demonstrated on two examples from quantum chemistry and physics.
\begin{itemize}
\item[-] The Hamiltonian of the well-studied hydrogen molecule in minimal basis is re-shaped into a two-qubit problem, using a simple code.
\item[-] A Fermi-Hubbard model on a $2 \times 5$ lattice and  periodic boundary conditions in the lateral direction is considered. We parametrize and  compare the sizes of the resulting Hamiltonians, as we employ different codes to save various amounts of qubits. In this way, the trade-off between qubit savings and gate complexity is illustrated (see Table \ref{tab:onlytable}).
\end{itemize}
\end{itemize}
\newpage
\onecolumngrid

 \begin{table}
\begin{tabular}{l|c|c|c|c|c}
Mapping & En-/Decoding type & Qubits saved & $n(N,K)$ & Resulting gates & Origin \\  \hline
Jordan-Wigner $|$ Parity transform & linear/linear & none & $N$ & length-$O(n)$ Pauli strings & \cite{wigner1928uber, seeley2012bravyi} \\
Bravyi-Kitaev transform & linear/linear & none & $N$ & length-$O(\log n)$ Pauli strings& \cite{bravyi2002fermionic}\\
\Parties & linear/ affine linear& $O(1)$ & $N-1$ & length-$O(n)$ Pauli strings& here\\
Binary addressing codes & non-linear/non-linear & $O( 2^{n/K})$ & $\log\left( N^K/K! \right)$ & $\left(O(n)\right)$-controlled gates & here \\
Segment codes & linear/non-linear&  $O\left(n/K\right)$ & $ N/(1+\frac{1}{2K})$ & $\left(O(K) \right)$-controlled gates & here
\end{tabular}
\caption{ Overview of mappings presented in this paper, listed by the complexity of their code functions,  their qubit savings, qubit requirements ($n$), properties of the resulting gates and first appearance. Mappings can be compared with respect to the size of plain words ($N$) and their targeted Hamming weight $K$. We also refer to different methods that are not listed, as they do not rely on codes in any way \cite{moll2016optimizing, bravyi2017tapering}.  }
\label{tab:othertable}
\end{table}
\twocolumngrid

\section{Background}
\label{sec:background}
To illustrate the general use of (possibly non-linear) encodings to represent fermionic models, let us first briefly generalize how existing mappings can be phrased in terms of linear encodings in the spirit of~\cite{seeley2012bravyi}.
Under consideration in representing the dynamics is a mapping for
second-quantized Hamiltonians of the form
\begin{align}
H&= \sum_{l=0}^{\infty}\; \sum_{\substack{\boldsymbol{a}\in[N]^{\otimes l} \\ \,\bbs{b}\in \mathbb{Z}_2^{\otimes l}}} h_{\bbs{ab}}\, \prod_{i=1}^l (c^{\dagger}_{a_i} )^{b_i}(c^{\,}_{a_i})^{1+b_i\moto} \nonumber \\ \label{eq:generichamiltonian}&= \sum_{l}\sum_{\substack{\bbs{a}, \bbs{b} \\\,\text{with}\; h_{\bbs{ab}}\neq 0}} \hat{h}_{\bbs{ab}}\, ,
\end{align}
where $h_{\bbs{ab}} $ are complex coefficients, chosen in a way as to render $H$ Hermitian.   We illustrate the use of such a mapping in the context of quantum simulation in Appendix \ref{sec:hamilsim}.
For our convenience, we use length-$l$ $N$-ary vectors $\bbs{a}=(a_1, \, ... \, , a_l)^\top \in [N]^{\otimes l}$ to parametrize the orbitals on which a term $\hat{h}_{\bbs{a}\,\bbs{b}}$ is acting, and write $[N] = \lbrace 1, \, ... \, , N \rbrace$. A similar notation will be employed for binary vectors of length $l$, with $\bbs{b}=(b_1, \, ... \, , b_l)^\top\in\mathbb{Z}_2^{\otimes l}$, $\mathbb{Z}_2=\lbrace 0, 1 \rbrace$, deciding whether an operator is a creator or annihilator by the rules $(c^{(\dagger)}_i)^1=c^{(\dagger)}_i$ and $(c^{(\dagger)}_i)^0=1$. \\
Every term $\hat{h}_{\bbs{ab}}$ is a linear operation $\mathcal{F}_N\to \mathcal{F}_N$, with $\mathcal{F}_N$ being the Fock space restricted on $N$ orbitals, the direct sum of all possible anti-symmetrized $M$-particle Hilbert spaces $\mathcal{H}_N^M$: $\mathcal{F}_N=\bigoplus_{m=0}^N \mathcal{H}_N^{m}$. Conventional mappings transform states of the Fock space $\mathcal{F}_N$ into states on $N$ qubits, carrying over all linear operations as well $\mathcal{L}(\mathcal{F}_N) \to \mathcal{L}((\mathbb{C}^2)^{\otimes N})$. \\
Before we start presenting conventional transformation schemes, we need to make a few  remarks on transformed Hamiltonians and notations pertaining to them. First of all, we identify the set of gates $\left\lbrace  \mathcal{P}, \mathbb{I} \right\rbrace^{\otimes n} = \left\lbrace  X, Y , Z, \mathbb{I} \right\rbrace^{\otimes n}$ with the term Pauli strings (on $n$ qubits).
The previously mentioned Jordan-Wigner transform, obviously has the power to transform  \eqref{eq:generichamiltonian} into a Hamiltonian that is a weighted sum of Pauli strings on $N$ qubits. General transforms, however, might involve other types of gates. We however have the choice to decompose these into Pauli strings. One might want to do so when using standard techniques for Hamiltonian simulation.
 In the following, we will denote the correspondence of second quantized operators or states $B$ to their qubit counterparts $C$ by: $B\; \hat{=}\;C$. For convenience, we will also omit identities in Pauli strings and rather introduce qubit labels, e.g. $X\otimes\mathbb{I} \otimes X=X_1\otimes X_3 =(\bigotimes_{i\in\lbrace 1,3\rbrace}X_i)$ and write $\mathbb{I}^{\otimes n}=\mathbb{I}$.

Consider a linear encoding of $N$ fermionic sites into $n=N$ qubits given by a binary matrix $A$ such that
\begin{align}
\label{eq:conventional}
\ket{\bbs{\omega}} = \ket{\enc\left(\bbs{\nu}\right)} =  \ket{A\bbs{\nu}\moto}\hat{=} \left(\prod_{j=1}^{N}(c_j^\dagger)^{\nu_j}\right)\ket{\Theta}
\end{align}
and $A$ is invertible, i.e. $\left(A A^{-1} \moto\right) = \id$. Note that in this case, the decoding given by $~\bbs{\nu} = ~\dec( \bbs{\omega}) =~\left({A^{-1}\bbs{\omega}\moto}\right)$ is also linear. It is known that any such matrix $A$, subsequently also yields a mapping of the fermionic creation and annihilation operators to qubit gates~\cite{seeley2012bravyi}.
To see how these are constructed, let us start by noting that they must fulfill the properties given in~\eqref{eq:stateaction1}-\eqref{eq:stateaction4} and~\eqref{eq:antiraw}, which motivates the definition of a parity, a flip and an update set below:
\begin{enumerate}
\item $c_{i_m}^{(\dagger)}$ anticommutes with the first $m-1$ operators and thus acquires phase $(-1)^{m-1}$.
\item A creation operator $c_{i_m}^\dagger$ might be absent (present) in between   $c^\dagger_{i_{m-1}}$ and $c^\dagger_{i_{m+1}}$, leading the rightmost operator  $c_{i_m}^{(\dagger)}$ to map the entire state to zero since  $c_{i_m}\ket{\Theta}=0$ $\left( c_{i_m}^\dagger c_{i_m}^\dagger=0 \right)$.
\item Given that the state was not annihilated, the occupation of site $i_m$ has to be changed. This means a creation operator $c_{i_m}^\dagger$ has to be added or removed between $c^\dagger_{i_{m-1}}$ and $c^\dagger_{i_{m+1}}$.
\end{enumerate}
These rules tell us what the transform of an operator $c_j^{(\dagger)}$ has to inflict on a basis state \eqref{eq:conventional}.
In order to implement the phase shift of the first rule,  a series of  Pauli-$Z$ operators is applied on qubits, whose numbers are in the \textit{parity set} (with respect to $j\in [N]$), $P(j) \subseteq [N]$.
Following the second rule we project onto the $\pm 1$ subspace of the $Z$-string on qubits indexed by another $[N]$ subset, the so-called \textit{flip set} of $j$, $F(j)$. The \textit{update set} of $j$, $U(j)\subseteq [N]$ labels the qubits to be flipped completing the third rule using an $X$-string.
\begin{align}
\label{eq:line}
&(c^{\dagger}_j)^b ( c^{\,}_j )^{b+1 \moto}\;\hat{=}\;\nonumber \\&\frac{1}{2}\left(\bigotimes_{k\in U(j)} X_k\right)\left(\mathbb{I}-(-1)^b \bigotimes_{l\in F(j)} Z_l\right)  \bigotimes_{m\in P(j)} Z_m \, ,
\end{align}
with $b\in\mathbb{Z}_2$. $P(j)$, $F(j)$ and $U(j)$  depend on the matrices $A$ and $A^{-1}$  as well as the parity matrix $R$. The latter is a $(N\times N)$ binary matrix which has its lower triangle filled with ones, but not  its diagonal. For the matrix entries this means $R_{ij}= \theta_{ij}$,  with $\theta_{ij}$ as the discrete version of the Heaviside function  \begin{align}
\label{eq:heavy}
\theta_{ij}=\begin{cases}0& i \leq j \\ 1 & i>j  \; . \end{cases}
\end{align}
The set members are obtained in the following fashion:
\begin{enumerate}
\item $P(j)$ contains all column numbers in which the $j$-th row of matrix $(R A^{-1} \moto)$ has non-zero entries.
\item $F(j)$ contains  the column labels of non-zero entries in the $j$-th row of $A^{-1}$.
\item $U(j)$ contains all row numbers in which the $j$-th column of $A$ has non-zero entries.
\end{enumerate}

Note that this definition of the sets differs from their original appearance in \cite{seeley2012bravyi,tranter2015bravyi}, where diagonal elements are not included. In this way, our sets are not disjoint, which leads to $Z$-cancellations and appearance of Pauli-$Y$ operators, but we have generalized the sets for arbitrary invertible matrices, and provided a pattern for other transforms later.  In fact, we recover these linear transforms from the general case in Appendix \ref{sec:linear}. There we also show explicitly that these operators abide by  \eqref{eq:stateaction1}-\eqref{eq:antiraw}.
\subsection{Jordan-Wigner, Parity and Bravyi-Kitaev transform}
As an illustration, we present popular examples of these linear transformations,  note again that all of these will have $n=N$.
The Jordan-Wigner transform is a special case for $A = \id$, leading to the direct mapping.
The operator transform gives $L= O(N)$ Pauli strings as
\begin{align}
&(c^{\dagger}_j)^b ( c^{\,}_j )^{b+1 \moto}\;\hat{=}\;\frac{1}{2}\left(X_j+i(-1)^b \, Y_j\right)  \bigotimes_{m<j} Z_m \, .
\end{align}
In the parity transform~\cite{seeley2012bravyi}, we have $L=O(N)$ $X$-strings:
\begin{align}
A^{-1}=\left[\begin{matrix}
1  \\
1 & 1 \\
 & \ddots	&  \ddots	\\
 & & 1 & 1
	 \end{matrix} \right] \;, \quad A=\left[\begin{matrix}
1  \\
1& 1 \\
\vdots & \vdots	&  \ddots	\\
 1&1 & \cdots & 1
	 \end{matrix} \right] \; ,
	 \end{align} \begin{align} \nonumber
&(c_j^{\dagger})^{b}(c_j^{\,})^{b+1 \moto}\; \hat{=}\; \\ &\frac{1}{2} \left( Z_{j-1}\otimes X_j + i(-1)^b\,  Y_j\right)\bigotimes_{m=j+1}^{N} X_m \, .
\end{align}
The Bravyi-Kitaev transform~\cite{bravyi2002fermionic} is  defined by a matrix $A$ \cite{seeley2012bravyi,tranter2015bravyi} that has non-zero entries according to a certain binary tree rule, achieving $L=O(\log N)$.

\subsection{Saving qubits by exploiting symmetries}
Our goal is to be able to trade quantum resources, which is done by reducing  degrees of freedom by exploiting symmetries. For that purpose, we provide a theoretical foundation to characterize the latter. \\
Parity, Jordan-Wigner and Bravyi-Kitaev transforms  encode all $\mathcal{F}_N$ states and  provide mappings for every $\mathcal{L}\left( \mathcal{F}_N \right)$ operator. Unfortunately, they require us to own a  $N$-qubit quantum computer, which might be unnecessary.
In fact, the only operator we want to simulate is the Hamiltonian, which usually has certain symmetries. Taking these symmetries into account  enables us to perform the same task with $n\leq N$ qubits instead. Symmetries usually divide the $\mathcal{F}_N$ into subspaces, and the idea is to encode only one of those.  Let $\mathcal{B}$ be a basis spanning a subspace $\mathrm{span}(\mathcal{B})\subseteq \mathcal{F}_N$   be associated with a Hamiltonian \eqref{eq:generichamiltonian}, where for every $l$, $\bbs{a}$, $\bbs{b}$; $\hat{h}_{\bbs{ab}}: \mathrm{span}(\mathcal{B}) \to \mathrm{span}(\mathcal{B})$. Usually, Hamiltonian symmetries generate many such (distinct) subspaces. Under consideration of additional information about our problem, like particle number, parity or spin polarization, we select the correct subspace. Note that particle number conservation is by far the most prominent symmetry to take into account. It is generated by Hamiltonians that are linear combinations of products of $ c^{\dagger}_ic^{\,}_j\, | \,i,j\in [N] $. These Hamiltonians, originating from first principles, only exhibit terms conserving the total particle number; $\hat{h}_{\bbs{ab}}: \mathcal{H}_N^M \to \mathcal{H}_N^M$. From all the Hilbert spaces $\mathcal{H}_N^M$, one considers the space with the particle number matching the problem description. \\ These symmetries will be utilized  in the next section: we develop a language that allows for encodings $\bbs{e}$ that reduce the length of the binary vectors $\bbs{e}( \bbs{\nu} )$ as compared to $\bbs{\nu}$.  This means that the state $\bbs{\nu}$ will be encoded in $n\leq N$ qubits, since each digit saved corresponds to a qubit eliminated.  As suggested by Bravyi et al.~\cite{bravyi2017tapering}, qubit savings can be achieved under the consideration of non-square, invertible matrices  $A$.  However, we will see below that using transformations based on non-linear encodings and decodings $\bbs{d}$ (the inverse transform defined by $A^{-1}$ before), we can eliminate a number of qubits that scales with the system size. For linear codes on the other hand, we find a mere constant saving.
\section{General transformations}
We here show how second-quantized operators and states,  Hamiltonian symmetries and the fermionic basis $\mathcal{B}$ are fused into a simple description  of  occupation basis states. While in this section all general ideas are presented, we would like to refer the reader to the appendices for details: to Appendix \ref{sec:opmap} in particular, which holds the proof of the underlying techniques.  Fermionic basis states are  represented by binary vectors $\bbs{\nu} \in \mathbb{Z}_2^{\otimes N}$, with its components implicating the occupation of the corresponding orbitals. Basis states inside the quantum computer, on the other hand, are represented  by binary vectors on a smaller space $\bbs{\omega}\in \mathbb{Z}_2^{\otimes n}$. These vectors  are code words of the former $\bbs{\nu}$,  where the binary code connecting all $\bbs{\nu}$ and $\bbs{\omega}$ is possibly non-linear.  In the end, an instance of such a code will be sufficient to describe states and operators, in a similar way than the matrix pair ($A$,~$A^{-1}$) governs the conventional transforms already presented. We now start by defining such codes and connect them to the state mappings. \\
Let $\mathrm{span}\left(\mathcal{B} \right)$ be a subspace of $\mathcal{F}_N$, as defined previously. For $n \geq \log |\mathcal{B}|$, we define two binary vector functions $\bbs{d}:\mathbb{Z}_2^{\otimes n} \to \mathbb{Z}_2^{\otimes N} $, $\bbs{e}:\mathbb{Z}_2^{\otimes N} \to \mathbb{Z}_2^{\otimes n} $, where we regard each  component as a binary function  $\bbs{d}=(d_1, \, ...\, , d_N)\,|\, d_i:\mathbb{Z}_2^{\otimes n} \to \mathbb{Z}_2$. Furthermore we introduce the binary basis set $\mathcal{V}\subseteq \mathbb{Z}_2^{\otimes N} $, with
\begin{align}
\label{eq:btov}
\bbs{\nu}\in\mathcal{V}, \quad \text{only if} \quad \left(\prod_{i=1}^N \left( c^\dagger_i \right)^{\nu_i} \right) \ket{\Theta} \in \mathcal{B} \, .
\end{align}
All elements in $\mathcal{B}$  shall be represented in $\mathcal{V}$. If for all $\bbs{\nu}\in \mathcal{V}$ the binary functions $\bbs{e}$ and $\bbs{d}$ satisfy $\bbs{d}\left( \bbs{e}\left( \bbs{\nu}\right)\right)=\bbs{\nu}$, and for all $\bbs{\omega}\in\mathbb{Z}_2^{\otimes n}: \,\bbs{d}\left( \bbs{\omega} \right) \in \mathcal{V}$, then we call the two functions encoding and decoding, respectively. An encoding-decoding pair  ($\bbs{e}$, $\bbs{d}$) forms a code.   \\ We thus have obtained a general form of  encoding, in which  qubit states only represent the subspace $\mathrm{span}\left(\mathcal{B}\right)$.  The decoding, on the other hand,  translates the qubit basis back to the fermionic one:
\begin{align}
\label{eq:statemap}
 \ket{\boldsymbol{\omega}} := \bigotimes_{j=1}^n \ket{\omega_j}  \; \hat{=}\; \left(\prod_{i=1}^{N} ( c_i^\dagger)^{d_i\left(\boldsymbol{\omega}  \right)} \right) \ket{\Theta} \, .
\end{align}
We intentionally keep the description of these functions abstract, as the  code used might be non-linear, i.e. it cannot be described with matrices $A$, $A^{-1}$. Non-linearity is thereby predominantly  encountered in  decoding rather than in encoding functions, as we will see in the examples obtained later. \\ For any code ($\bbs{e}$, $\bbs{d}$), we will now present the transform of fermionic operators into qubit gates.
Before we can do so however, two issues are to be addressed.
Firstly, one observes that we cannot hope to find a transformation recipe for a singular fermionic operator $c_j^{(\dagger)}$. The reason for this is that the latter operator changes the occupation of the $j$-th orbital. As a consequence, a state with the occupation vector $\bbs{\nu}$ is mapped to $\left(\bbs{\nu}+\bbs{u_j}\moto \right)$, where $\bbs{u_j}$ is the unit vector of component $j$; $\left(u_j\right)_i=\delta_{ij}$. The problem is that since we have trimmed the basis, $(\bbs{\nu}+\bbs{u_j}\moto)$ will probably not be in $\mathcal{V}$, which means this state is not encoded ~\footnote{`Unencoded  state' is  actually a slightly misleading term: when we say a state $\bbs{\lambda}\in \mathbb{Z}_2^{\otimes N}$ is not encoded, we actually mean that it cannot be encoded and correctly decoded, so $\bbs{d}\left(\bbs{e}\left(\bbs{\lambda}\right)\right)\neq \bbs{\lambda}$.}. The action of $c_j^{(\dagger)}$ is, thus, not defined.  We can however obtain a recipe for the non-vanishing Hamiltonian terms $\hat{h}_{\bbs{a}\bbs{b}}$ as they do not escape the encoded space being  $\left(\mathrm{span}(\mathcal{B})\to \mathrm{span}(\mathcal{B})\right)$-operators. Note that this issue is never encountered in the conventional transforms, as they encode the  entire Fock space.\\
Secondly, we are yet to introduce a tool to transform fermionic operators into quantum gates. The structure of the latter has to be similar to the linear case, as they mimic the same dynamics as presented in Section \ref{sec:background}.  In general, a gate sequence will commence with some kind of projectors into the subspace with the correct occupation,  as well as operators implementing parity phase shifts. The sequence should  close with bit flips to update the state.  The task is now to determine the form of these operators. The issue boils down  to finding operators that extract binary information from qubit states, and map it onto their phase.  In other words, we need to find  linear operators associated with e.g.  the binary function $d_j$,  such that it maps basis states $\ket{\bbs{\omega}}\to(-1)^{d_j(\bbs{\omega})}\ket{\bbs{\omega}}$. In any case, we must recover the case of Pauli strings on their respective sets when considering linear codes. For our example, this means the linear case yields the operator $(\bigotimes_{m\in F(j)} Z_m )$. Using general codes,  we  are lead to define the extraction superoperation $\mathfrak{X}$, which maps binary functions to quantum gates on $n$ qubits:
\begin{align}
\mathfrak{X}: \left(\mathbb{Z}_2^{\otimes n} \to \mathbb{Z}_2 \right) \to \mathcal{L}\left( (\mathbb{C}^2)^{\otimes n}\right) \, .
\end{align}
 The extraction superoperator is defined for all binary vectors $\bbs{\omega}\in \mathbb{Z}_2^{\otimes n}$ and binary functions $f,g:\mathbb{Z}_2^{\otimes n} \to \mathbb{Z}_2$ as:

\begin{align}
 &\mathfrak{X}[f] \ket{\bbs{\omega}} = (-1)^{f(\bbs{\omega})}  \ket{\bbs{\omega}}
\nonumber \\\label{eq:prop1} &\text{(Extraction property)} \end{align} \begin{align}
\mathfrak{X}\left[\bbs{\omega}\to\right. & \left. f\left( \bbs{\omega}\right)+g \left(\bbs{\omega} \right) \moto\right]= \mathfrak{X}[f]\;\mathfrak{X}[g]  \nonumber \\ \label{eq:prop2} &\text{(Exponentiation identity)}\end{align} \begin{align}
& \; \; \mathfrak{X}\left[ \bbs{\omega}\to b\right]=(-1)^b \,\mathbb{I}\quad |\; b\in \mathbb{Z}_2 \nonumber \\ \label{eq:prop3}& \text{(Extracting constant functions)}\end{align} \begin{align}
&\; \; \mathfrak{X}\left[\bbs{\omega} \to \omega_j \right] = Z_j \quad | \; j \in [n] \nonumber \\ \label{eq:prop4}& \text{(Extracting linear functions)} \end{align} \begin{align}
& \mathfrak{X}\left[\bbs{\omega} \to \prod_{j\in \mathcal{S}}\omega_{j}\right] = \mathrm{C}^{k}\,\mathrm{\textsc{Phase}}(i_1, ...\, , i_{k+1})
 \nonumber \\ &\;\;\;\;\;\;\text{with} \;\; \mathcal{S}=\lbrace i_s\rbrace_{s=1}^{k+1} \subseteq [n], \;\; k\in [n-1] \nonumber \\ \label{eq:prop5}& \;\;\;\;\;\;\text{(Extracting non-linear functions).}
\end{align}
Note that the first two properties imply that the operators $\mathfrak{X}[f]$, $\mathfrak{X}[g] $ commute and all operators are diagonal in the computational basis. Given that binary functions have a  polynomial form, we are now able to construct operators by extracting every binary function possible, for example
\begin{align}
&\mathfrak{X}[\bbs{\omega}\to 1+\omega_1+\omega_1\omega_2\moto]\nonumber \\ \label{eq:ie1}
  &= \mathfrak{X}\left[ \bbs{\omega}\to1 \right] \; \mathfrak{X}\left[ \bbs{\omega}\to \omega_1 \right] \; \mathfrak{X}\left[ \bbs{\omega}\to \omega_1 \omega_2 \right]\\ \label{eq:ie2}
&= -Z_1\,\mathrm{\textsc{CPhase}}(1,2) \,.
\end{align}
We firstly we have used \eqref{eq:prop2} to arrive at \eqref{eq:ie1}, and then reach \eqref{eq:ie2} by applying the properties  \eqref{eq:prop3}-\eqref{eq:prop5} to the respective sub-terms. This might however not be the final Hamiltonian, since the simulation algorithm  might require us to reformulate the Hamiltonian as a sum of weighted  Pauli strings \cite{whitfield2011simulation,hastings2014improving}. In that case, need to decompose all controlled gates. The cost for this decomposition is  an increase in the number of Hamiltonian terms, for instance we find $\mathrm{\textsc{CPhase}}(i,j)=\frac{1}{2}(\mathbb{I}+Z_i+Z_j-Z_i  \otimes Z_j)$. In general, \eqref{eq:prop4} and \eqref{eq:prop5} can be replaced by an adjusted definition:
 \begin{align}
 &\mathfrak{X}\left[\bbs{\omega}\to \prod_{j\,\in\, \mathcal{S}} \omega_j \right] = \left. \mathbb{I}-2\prod_{j \,\in\, \mathcal{S}} \frac{1}{2} \left(\mathbb{I}-Z_j \right) \quad \right|\; \mathcal{S}\subseteq [n] \nonumber \\ \label{eq:prop6} &\text{(extracting non-constant functions).}
\end{align}
We will be able to define the operator mappings introducing the parity and update functions, $\bbs{p}$ and $\bbs{\varepsilon}^{\,\bbs{q}}$:
\begin{align}  \label{eq:p}
 &\bbs{p}:\mathbb{Z}_2^{\otimes n}\to\mathbb{Z}_2^{\otimes N}, \quad p_j\left(\bbs{\omega} \right) = \sum_{i=1}^{j-1} d_i\left(\bbs{\omega} \right)\moto \, , \\
 &\bbs{\varepsilon}^{\, \bbs{q}}:\mathbb{Z}_2^{\otimes n}\to\mathbb{Z}_2^{\otimes n}, \quad \text{with}\; \bbs{q}\in \mathbb{Z}_2^{\otimes N}\nonumber \\ \label{eq:varepsilon}  &\bbs{\varepsilon}^{\, \bbs{q}}\left(\bbs{\omega} \right) = \bbs{e}\left( \bbs{d}\left(    \bbs{\omega}\right)+\bbs{q}\;\moto \right)+ \bbs{\omega} \;\moto \, .
\end{align}
Finally, we have collected all the means to obtain the operator mapping for weight-$l$ operator sequences as they occur in \eqref{eq:generichamiltonian}:
\begin{align}
\label{eq:admiral}
&\prod_{i=1}^l (c^{\dagger}_{a_i} )^{b_i}(c^{\,}_{a_i})^{1+b_i\moto}\; \hat{=}\; \;\mathcal{U}^{\,\bbs{a}} \left( \prod_{v=1}^{l-1}\,\prod_{w=v+1}^{l}\left(-1 \right)^{\theta_{a_v a_w}} \right)\nonumber \\  &\times   \prod_{x=1}^{l} \frac{1}{2}\left(\mathbb{I}- \left[\prod_{y=x+1}^{l}(-1)^{\delta_{a_x a_y}}\right](-1)^{b_x}\;\mathfrak{X}\left[d_{a_x}\right]\right) \mathfrak{X}\left[ \,p_{a_x} \right]
\end{align}
where $\theta_{ij}$ is defined in \eqref{eq:heavy} and $\delta_{ij}$ is the Kronecker delta. In this expression, we find various projectors, parity operators with corrections for occupations that have changed before the update operator is applied.
The update operator $\mathcal{U}^{\, \bbs{a}}$, is characterized by the $\mathbb{Z}_2^{\otimes N}$-vector $\bbs{q}=\sum_{i=1}^l \bbs{u_{a_i}} \moto$.
\begin{align}
\label{eq:updateop}
\mathcal{U}^{\,\bbs{a}} = \sum_{\bbs{t}\, \in\, \mathbb{Z}_2^{\otimes n}}\left[\bigotimes_{i=1}^{n} \left(X_i\right)^{t_i}\right] \prod_{j=1}^{n} \frac{1}{2}\left( \mathbb{I}+(-1)^{t_j}\;\mathfrak{X}\left[ \varepsilon^{\, \bbs{q}}_j \right]\right).
\end{align}
This is a problem: when summing over the  entire  $\mathbb{Z}_2^{\otimes n}$, one has to expect an exponential number of terms. As a remedy, one can arrange the resulting operations into controlled gates, or rely on codes with a linear encoding.  If the encoding can be defined using a binary  $(n\times N)$-matrix $A$,  $\bbs{e}\left( \bbs{\nu}\right)= \left(A\bbs{\nu}\moto \right)$, the update operator reduces to
\begin{align}
\label{eq:linenc}
\mathcal{U}^{\,\bbs{a}} = \bigotimes_{i=1}^{n} \left(X_i\right)^{\sum_{j}A_{ij} q_j \moto} \, .
\end{align}
 In Appendix \ref{sec:opmap}, we show that \eqref{eq:admiral}-\eqref{eq:linenc} satisfy the conditions \eqref{eq:stateaction1}-\eqref{eq:antiraw}. In the following we will turn our attention to the most fruitful symmetry to take into account:  particle conservation symmetry. While code families accounting for this symmetry are explored in the next subsection, alternatives to the mapping of entire Hamiltonian terms are discussed for such codes in Appendix \ref{sec:altmaps}.
\subsection{Particle number conserving codes}
In the following, we will present three types of codes that save qubits by exploiting particle number conservation symmetry, and possibly the conservation of the total spin polarization.
Particle number conserving Hamiltonians are highly relevant for quantum chemistry and problems posed from first principles.  We  therefore set out to find codes in which $\bbs{\nu} \in \mathcal{V} $ have a constant Hamming weight $\hamming{\bbs{\nu}}=K$. Since the Hamming weight is defined as $\hamming{\bbs{\nu}}=\sum_{m} \nu_m$, it yields the total occupation number for the vectors $\bbs{\nu}$. In order to simulate systems with a fixed particle number, we are thus interested to find codes that implement code words of constant Hamming weight.    Note that the fixed Hamming weight $K$ does not necessarily need to coincide with the total particle number $M$. A code with the latter property might also be interesting for systems with additional symmetries. Most importantly, we have not taken into account the spin multiplicity yet. As the particles in our system are fermions, every spatial site will typically have an even number of spin configurations associated with it. Orbitals with the same spin configurations naturally denote subsets of the total amount of orbitals, much like the suits in a card deck.  An absence of magnetic terms as well as spin-orbit interactions leaves the Hamiltonian to conserve the number of particles inside all those suits. Consequently, we can append several constant-weight codes to each other. Each of those subcodes encodes thereby  the orbitals inside one suit. In electronic system with only Coulomb interactions for instance, we can  use  two subcodes ($\bbs{e^{\,\diamondsuit}}$,~$\bbs{d^{\,\diamondsuit}}$) and ($\bbs{e^{\,\spadesuit}}$,~$\bbs{d^{\,\spadesuit}}$), to encode  all spin-up, and  spin-down orbitals, respectively. The global code ($\bbs{e}$, $\bbs{d}$), encoding the entire system,  is obtained by appending the subcode functions e.g. $\bbs{d}\left(\bbs{\omega^1}\oplus \bbs{\omega^2}\right)=\bbs{d^{\,\diamondsuit}}(\bbs{\omega^{1}})\oplus \bbs{d^{\,\spadesuit}}(\bbs{\omega^{2}}).$ Appending codes like this will help us to achieve higher savings at a lower gate cost.  \\ The codes that we now introduce (see also again Table \ref{tab:othertable}), fulfill the task of encoding only constant-weight words differently well. The larger $\mathcal{V}$, the less qubits will be eliminated, but we expect the resulting gate sequences to be more simple. Although not just words of  that weight are encoded, we treat $K$ as a parameter - the targeted weight.
\subsubsection{\Parties }
\label{subsubsec:party}
A slim, constant amount of qubits can be saved with the following $n=N-1$, affine linear codes.  \Parties encode all the words with either even or odd Hamming weight.  As this corresponds to exactly half of the Fock space, one qubit is eliminated.  This means we disregard the last component when we encode $\bbs{\nu}$ into words with one digit less. The decoding function then adds the missing component depending on the parity of the code words. The code for $K$ odd is defined as
\begin{align} \bbs{d}\left( \boldsymbol{\omega}\right)&=
\left[\begin{matrix}
1 \\
& \ddots \\
& & 1 \\
1 & \cdots & 1
\end{matrix} \right]\bbs{\omega} + \left( \begin{matrix}
0 \\ \vdots  \\0 \\1
\end{matrix} \right) \mod 2 \, ,
\end{align}
\begin{align}
\bbs{e}\left( \boldsymbol{\nu}\right)&=
\left[\begin{matrix}
1  & & & 0\\
& \ddots & & \vdots \\
& & 1 & 0
\end{matrix} \right]\boldsymbol{\nu} \mod 2 \, .
\end{align}
In the even-$K$ version,  the affine vector $\bbs{u_N}$, added in the decoding, is removed. Since encoding and decoding function are both at most affine linear, the extracted operators will all be Pauli strings, with at most a minus sign. The advantage of the \parties is that they do not depend on $K$. They can be used even in cases of smaller saving opportunities, like $K\approx N/2$. We can employ these codes even for  Hamiltonians that conserve only the fermion parity. This makes them important for effective descriptions of superconductors \cite{ruckenstein1987mean}.
\subsubsection{Codes with binary addressing} \label{subsubsec:binadressing}
We present a concept for  heavily non-linear  codes for large qubit savings, $n=\lceil\log (N^K/K!)\rceil$, \cite{tian2007constant}. In order to conserve the maximum amount of qubits possible, we choose to encode particle coordinates as binary numbers in $\bbs{\omega}$. To keep it simple, we here consider the example of weight-one binary addressing codes, and refer the reader to  Appendix \ref{sec:binary} for $K>1$. In $K=1$, we recognize the qubit savings to be exponential, so consider $N=2^n$. Encoding and decoding functions are defined by means of the binary enumerator, $\mathrm{bin}:\mathbb{Z}_2^{\otimes n} \to \mathbb{Z}$, with $\mathrm{bin}\left( \bbs{\omega}\right)=\sum_{j=1}^{n} 2^{j-1} \omega_j $.
\begin{align}\label{eq:bindec}
d_j\left(\bbs{\omega} \right) = \prod_{i=1}^{n}\left( \omega_i + 1 +q_i^{\,j} \right) \moto\, , \end{align}
\begin{align}\label{eq:binenc}
\bbs{e}\left( \bbs{\nu} \right) = \left[\begin{tabular}{c|c|c|c}  & & &   \\ $\bbs{q^1}$ & $\bbs{q^2}$& $\quad \cdots \quad $ & $\bbs{q^{^{2^n}}}$  \\  & & &  \end{tabular} \right] \bbs{\nu} \;\moto \; ,
\end{align}
where $\bbs{q^{\,j}}\in \mathbb{Z}_2^{\otimes n}$ is implicitly defined by $\mathrm{bin}(\bbs{q^j})+1=j$. An input $\bbs{\omega}$ will by construction render only  the $j$-th component of \eqref{eq:bindec} non-zero, when $\bbs{q^{\, j}}=\bbs{\omega}$ \footnote{ For better or worse we have used the binary representation of the orbital indexes. We could however employ any other counting method, i.e. any injective mapping that relates a binary vector representing a qubit basis state to an integer labeling an orbital.}. \\The exponential qubit saving comes at a high cost: the product over each  component of $\bbs{\omega}$ implies multi-controlled gates on the entire register. This is likely to cause connectivity problems. Note that decomposing the controlled gates will in general be practically prohibited by the sheer amount of resulting terms.  On top of those drawbacks, we also expect the encoding function to be non-linear for $K>1$.
\subsubsection{Segment codes}
\label{subsubsec:mentcodes}
We introduce a type of scaleable  $n=\lceil N/(1+\frac{1}{2K}) \rceil$ codes  to eliminate a linear amount of qubits. The idea of segment codes is to cut the vectors $\bbs{\nu}$ into smaller, constant-size vectors $\bbs{\hat{\nu}^i}\in \mathbb{Z}_2^{\otimes \hat{N}}$, such that $\bbs{\nu}=\bigoplus_i \bbs{\hat{\nu}^i} $. Each such segment $\bbs{\hat{\nu}^i}$ is encoded by a subcode. Although we have introduced the concept already, this segmentation is independent from our treatment of spin `suits'.
In order to construct a weight-$K$ global code, we append several instances  of the same subcode. Each of these subcodes codes is defined on $\hat{n}$ qubits, encoding $\hat{N}=\hat{n}+1$ orbitals. We deliberately have chosen to only save one qubit per segment in order to keep the segment size $\hat{N}(K)$ small. \\
We now turn our attention to the construction of these segment codes. As shown in Appendix \ref{sec:ments}, the segment sizes can be set to $\hat{n}=2K$ and $\hat{N}=2K+1$. As the global code is supposed to encode all $\bbs{\nu}\in \mathbb{Z}_2^{\otimes N}$ with Hamming weight $K$, each segment must encode all vectors from Hamming weight zero up to weight $K$. In this way, we guarantee that the encoded space contains the relevant, weight-$K$ subspace. This construction follows from the idea that each block contains equal or less than $K$ particles, but might as well be empty.
For each segment, the following  de- and encoding functions are found for  $\bbs{\hat{\omega}}\in \mathbb{Z}_2^{\otimes \hat{n}}$, $\bbs{\hat{\nu}}\in \mathbb{Z}_2^{\otimes \hat{N}}$:
\begin{align}
\label{eq:segdec}
\bbs{\hat{d}}\left( \bbs{\hat{\omega}}\right)&=
\left[\begin{matrix}
1 \\
& \ddots \\
& & 1 \\
0 & \dots & 0
\end{matrix} \right]\bbs{\hat{\omega}} \;+\; f\left(\bbs{\hat{\omega}}\right)\left( \begin{matrix}
1 \\ \vdots \\ \vdots \\1
\end{matrix} \right) \moto
\end{align}
\begin{align}
\label{eq:segenc}
\bbs{\hat{e}}\left( \bbs{\hat{\nu}}\right)&=
\left[\begin{matrix}
1  & & &  1\\
& \ddots  & &  \vdots\\
& & 1  &  1
\end{matrix} \right]\bbs{\hat{\nu}} \mod 2 \, ,
\end{align}
where $f:\mathbb{Z}_2^{\otimes \hat{n}} \to \mathbb{Z}_2$ is a binary switch. The switch is the source of non-linearity in these codes. On an input $\bbs{\hat{\omega}}$ with $\hamming{\bbs{\hat{\omega}}}>K$, it yields one, and zero otherwise. \\
There is just one problem: segment codes are not suitable for particle-number conserving Hamiltonians, according to the definition of the basis $\mathcal{B}$, that we would have for segment codes.  The reason for this is that we have not encoded all states with $\hamming{\bbs{\nu}}>K $. In this way, Hamiltonian terms $\hat{h}_{\bbs{ab}}$ that exchange occupation numbers  between two segments, can map into unencoded space. We can, however, adjust these terms, such that they only act non-destructively on states  with at most $K$ particles between the involved segment. This does not change the model, but aligns the Hamiltonian with the necessary condition that we have on $\mathcal{B}$, $\hat{h}_{\bbs{a}\bbs{b}}: \mathrm{span}(\mathcal{B})\to \mathrm{span}(\mathcal{B})$.  This is discussed in detail Appendix \ref{sec:ments}, where we also provide  an explicit description of the binary switch mentioned earlier.

Using segment codes, the operator transforms will have multi-controlled gates as well: the binary switch is non-linear.  However, gates are controlled  on at most an entire segment, which means there is no gate that acts on more than $2K$ qubits. This an improvement in gate locality, as compared to  binary addressing codes.
\section{Examples}
\subsection{Hydrogen molecule}
In this subsection, we will demonstrate the Hamiltonian transformation on a simple problem. Choosing a standard example, we draw comparison with other methods for qubit reduction.
As one of the simplest problems, the minimal electronic structure of the hydrogen molecule has been studied extensively for quantum simulation \cite{aspuru2005simulated, whitfield2011simulation} already. We describe the system as two electrons on 2  spatial sites. Because of the spin-multiplicity, we require 4 qubits to simulate the Hamiltonian in conventional ways. Using the particle conservation symmetry of the Hamiltonian, this number can be reduced. The Hamiltonian also lacks terms that mix spin-up and -down states, with the  total spin polarization known to be zero. Taking into account these symmetries, one finds a total of 4 fermionic basis states: $\mathcal{V}=\left\lbrace \left(0,1,0,1\right),\left(0,1,1,0\right),\left(1,0,0,1\right),\left(1,0,1,0\right) \right\rbrace$. These can be encoded into two qubits by appending two instances of a ($N=2$, $n=1$, $K=1$)-code. The global code is defined as :
\begin{align}
\bbs{d}\left( \bbs{\omega} \right)&=\left[ \begin{matrix}
1 &   \\ 1 &  \\   & 1 \\   & 1
\end{matrix} \right] \bbs{\omega} +\left( \begin{matrix}
1 \\ 0 \\ 1 \\ 0
\end{matrix} \right) \moto \end{align}\begin{align} \bbs{e}\left( \bbs{\nu} \right)&= \left[ \begin{matrix}
0 & 1 & 0 & 0 \\
0 & 0 & 0 & 1
\end{matrix}\right] \bbs{\nu} \, \moto \, .
\end{align} The physical  Hamiltonian,
\begin{align}
\label{eq:hydrogenhamil}
H=\;&-h_{11}\left( c^{\dagger}_1c^{\;}_1+c^{\dagger}_3c^{\;}_3  \right)-h_{22}\left(c^{\dagger}_2c^{\;}_2  +c^{\dagger}_4c^{\;}_4 \right) \nonumber \\
&+ h_{1331}\;c^{\dagger}_1c^{\dagger}_3c^{\;}_3c^{\;}_1+ h_{2442}\;c^{\dagger}_2c^{\dagger}_4c^{\;}_4c^{\;}_2 \nonumber \\ &+h_{1221}\left( c^{\dagger}_1c^{\dagger}_4c^{\;}_4c^{\;}_1 + c^{\dagger}_3c^{\dagger}_2c^{\;}_2c^{\;}_3  \right) \nonumber \\ & +\left(h_{1221}-h_{1212}\right)\left(  c^{\dagger}_1c^{\dagger}_2c^{\;}_2c^{\;}_1+ c^{\dagger}_3c^{\dagger}_4c^{\;}_4c^{\;}_3  \right) \nonumber \\ &+h_{1212}\left( c^{\dagger}_1c^{\dagger}_4c^{\;}_3c^{\;}_2+c^{\dagger}_2c^{\dagger}_3c^{\;}_4c^{\;}_1    \right) \nonumber \\ &+h_{1212}\left(c^{\dagger}_1c^{\dagger}_3c^{\;}_4c^{\;}_2+c^{\dagger}_2c^{\dagger}_4c^{\;}_3c^{\;}_1 \right) \, ,
\end{align}
is  transformed into the qubit Hamiltonian \begin{align} g_{1}\; \mathbb{I} + g_{2}\; X_1\otimes X_2+g_{3} \;Z_1+g_{4}\; Z_2 + g_{5} \; Z_1 \otimes Z_2 \, .
\end{align}  The real coefficients $g_i$ are formed by the coefficients $h_{ijkl}$ of \eqref{eq:hydrogenhamil}.  The values of the $h_{ijkl}$ can be found in \citep{whitfield2011simulation}, for example. \\  In previous works, conventional  transforms have been applied to that problem Hamiltonian. Afterwards, the resulting 4-qubit-Hamiltonian has been reduced by hand in some way.
In \cite{o2016scalable}, the  actions on two  qubits are replaced with their expectation values after inspection of the Hamiltonian. In \cite{moll2016optimizing}, on the other hand, the Hamiltonian is reduced to two qubits in a systematic fashion. Finally, the case is revisited in  \cite{bravyi2017tapering}, where the problem is reduced below the combinatorical limit to one qubit. The latter two attempts have used Jordan-Wigner, the former the Bravyi-Kitaev transform first.

\subsection{Fermi-Hubbard model}
We present another example to illustrate the trade-off between qubit number and gate depth.
For that purpose, we consider a small lattice with periodic boundary conditions in the lateral direction. The system shall contain 10 spatial sites, doubled by the spin-multiplicity. The problem Hamiltonian is
\begin{align} \label{eq:FermiHubbardHamilton}
H=&-t\sum_{\left\langle i,j \right\rangle \in E} \left( c^{\dagger}_i c^{\;}_j+c^{\dagger}_jc^{\;}_i\right)\nonumber \\ &+U \sum_{j=1}^{10} c^{\dagger}_j c^{\;}_j\,c^{\dagger}_{10+j} c^{\;}_{10+j} \, ,
\end{align}
with its real coefficients $t,U$. It exhibits hopping terms along the edges $E$ of the graph in Figure  \ref{fig:fermihub}. The sketch on the left of this figure shows the connection graph of the first 10 orbitals. The other 10 orbitals are connected in the same fashion, and each such site is interacting with its counterpart from the other graph. We aim to populate this model with four fermions, where the total spin polarization is zero. Two conventional transforms and two transforms based on our codes are compared  by the amount of qubits necessary, as well as the size of the transformed Hamiltonian. As benchmarks, we decompose controlled gates and count the number of resulting Pauli strings. The sum of their total weight constitutes the gate count. Having these two disconnected graphs is an invitation to us to append two codes acting on sites $1-10$ and $11-20$ respectively. When using the $K=2$ segment code on one graph, the segments are formed as suggested by the right-hand side  of Figure \ref{fig:fermihub}. Note that from the combinatorical perspective, we could encode the problem with  11 qubits. However, if we append two $K=2$ binary addressing codes to each other,  the resulting Hamiltonian is on 14 qubits already. The problem is that the resulting Hamiltonian for this case cannot be expressed with decomposed controlled gates due to the high number of resulting terms.
Indeed, Table \ref{tab:onlytable} suggests that decomposing the controlling gates might easily lead to very large Hamiltonians with a multitude of very small terms.  The gate decomposition appears therefore undesirable. We in general recommend to rather decompose large controlled gates as shown in \cite{barenco1995elementary}. However, one also notices that an elimination of up to two qubits comes at a low cost: the amount of gates is not higher than in the Bravyi-Kitaev transform. As soon as we employ segment codes on the other hand, the Hamiltonian complexity rises with the amount of qubits eliminated.
\begin{figure}
\begin{tikzpicture}[scale=.6]
\node[] at (0,2) {};
\node[] at (0,-2) {};
\draw[] (0,0)--(4,0);
\draw[] (0,-1)--(4,-1);
\draw[] (0,0)--(0,-1);
\draw[] (1,0)--(1,-1);
\draw[] (2,0)--(2,-1);
\draw[] (3,0)--(3,-1);
\draw[] (4,0)--(4,-1);
\draw[] (-.8,0)--(0,0);
\draw[] (-.8,-1)--(0,-1);
\draw[] (4.8,0)--(4,0);
\draw[] (4.8,-1)--(4,-1);
\draw[color=black,fill=white] (0,0) circle[radius=0.2];
\draw[color=black,fill=white] (1,0) circle[radius=0.2];
\draw[color=black,fill=white] (2,0) circle[radius=0.2];
\draw[color=black,fill=white] (3,0) circle[radius=0.2];
\draw[color=black,fill=white] (4,0) circle[radius=0.2];
\draw[color=black,fill=white] (0,-1) circle[radius=0.2];
\draw[color=black,fill=white] (1,-1) circle[radius=0.2];
\draw[color=black,fill=white] (2,-1) circle[radius=0.2];
\draw[color=black,fill=white] (3,-1) circle[radius=0.2];
\draw[color=black,fill=white] (4,-1) circle[radius=0.2];
\node[below] at (0,-1.2) {1};
\node[below] at (1,-1.2) {2};
\node[below] at (2,-1.2) {3};
\node[below] at (3,-1.2) {4};
\node[below] at (4,-1.2) {5};
\node[above] at (0,0.2) {6};
\node[above] at (1,0.2) {7};
\node[above] at (2,0.2) {8};
\node[above] at (3,0.2) {9};
\node[above] at (4,0.2) {10};
 \end{tikzpicture} $\;\;$
\begin{tikzpicture}[scale=.6]
\node[] at (0,2) {};
\node[] at (0,-2) {};
\draw[] (0,0)--(4,0);
\draw[] (0,-1)--(4,-1);
\draw[ultra thick, color=black!50] (0,0)--(0,-1);
\draw[ultra thick, color=black!50] (1,0)--(1,-1);
\draw[ultra thick, color=black!50] (2,0)--(2,-1);
\draw[ultra thick, color=black!50] (3,0)--(3,-1);
\draw[ultra thick, color=black!50] (4,0)--(4,-1);
\draw[] (-.8,0)--(0,0);
\draw[] (-.8,-1)--(0,-1);
\draw[] (4.8,0)--(4,0);
\draw[] (4.8,-1)--(4,-1);
\draw[color=black,fill=white] (0,0) circle[radius=0.2];
\draw[color=black,fill=white] (1,0) circle[radius=0.2];
\draw[color=black,fill=white] (2,0) circle[radius=0.2];
\draw[color=black,fill=white] (3,0) circle[radius=0.2];
\draw[color=black,fill=white] (4,0) circle[radius=0.2];
\draw[color=black,fill=white] (0,-1) circle[radius=0.2];
\draw[color=black,fill=white] (1,-1) circle[radius=0.2];
\draw[color=black,fill=white] (2,-1) circle[radius=0.2];
\draw[color=black,fill=white] (3,-1) circle[radius=0.2];
\draw[color=black,fill=white] (4,-1) circle[radius=0.2];
\draw[ultra thick] (-.4,.3)--(4.4,.3)--(4.4,-.3)--(-.4,-.3)--cycle;
\draw[ultra thick] (-.4,-.7)--(4.4,-.7)--(4.4,-1.3)--(-.4,-1.3)--cycle;
\end{tikzpicture}
\caption{Left: illustration of the Fermi-Hubbard model considered. Lines between two sites, like 1 and 2, indicate the appearance of the term $t (c^{\dagger}_1c^{\;}_2 + c^{\dagger}_2c^{\;}_1)$ in the Hamiltonian \eqref{eq:FermiHubbardHamilton}. Periodic boundary conditions link sites 1 and 5 as well as 6 and 10. Sites 11-20 follow the same graph. Right: segmenting of the system; the two blocks are infringed. The gray links are  to be adjusted. }
\label{fig:fermihub}
\end{figure}
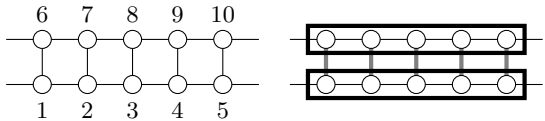
\begin{table}[h]
\begin{tabular}{l|c|c|c}
Mapping & Qubits & Gates & Terms  \\ \hline
Jordan-Wigner transform  & 20 & 232 & 74 \\
Bravyi-Kitaev transform & 20 & 278 & 74 \\
\Party$\oplus$ \Party  & 18 & 260 & 74 \\
\Party $\oplus$ Segment code & 17 & 4425 & 876 \\
Segment code $\oplus$ Segment code & 16 & 9366 & 1838
\end{tabular}
\caption{Relaxing the qubit requirements for the Hamiltonian \eqref{eq:FermiHubbardHamilton}, where various mappings trade  different amounts of qubits. The notation  $\oplus$ is used as two codes for different graphs are appended. We compare different mappings by the amount of qubits. We make comparisons by the number of Hamiltonian terms and the total weight of the resulting Pauli strings.  } \label{tab:onlytable}
\end{table}
\section{Conclusion and future work}
In this work, we have introduced new methods to reduce the number of qubits required for simulating fermionic systems in second quantization.
We see the virtue of the introduced concepts in the fact that it takes into account symmetries on a simple but non-abstract level. We merely concern ourselves with objects as simple as binary vectors, but  attribute the physical interpretation of orbital occupations to them. At this level, the mentioned symmetries are easy to apply and exploit.  The accounting for the complicated antisymmetrization of the many-body wave function  on the other hand is done in the fermionic operators, which to transform we have provided recipes for.  In these operator transforms we see room for improvement: we for instance lack a proper gate composition for update operators of non-linear encodings at this point. We on the other hand have the extraction superoperator $\mathfrak{X}$ return only conventional (multi)-controlled phase gates. Nonlinear codes would on the other hand benefit from a gate set  that includes gates with negative control, i.e. with the $(-1)$ eigenvalue conditioned on $\ket{0}$ eigenspaces of certain qubits involved.
We consider our work to be  relevant for quantum simulation with near-term devices, with a limited number of qubits at disposal. Remarks about asymptotic scaling are thus missing in this work, but would be interesting.    Also, we have centered our investigations around quantum computers with qubits. The idea behind the generalized operator transforms, however, can possibly be adapted to multi-level systems (qudits). The operator transforms of segment and binary addressing codes, for instance, might simplify in such a setup, if generalized Pauli operators are available in some form.   \\
Apart from the codes presented, we have laid the foundation for the reader to invent their own. As supplementary material, we include a program to transform arbitrary Hamiltonians from a second-quantized form into Pauli-string form, using user-defined codes. In this way we hope that in the long term, many more entries will be added to Table \ref{tab:othertable}.  Furthermore, we are certain that the table can be extended into another way:  gate relaxations for transforms with $n>N$ have already been shown \cite{bravyi2002fermionic,verstraete2005mapping,whitfield2016local,havlivcek2017operator}, and we are currently working in that direction.

\section{Acknowledgments}
We would like to thank Kenneth Goodenough, Victoria Lipinska, Thinh Le Phuc and Valentina Caprara Vivoli for helpful discussions on how to present our results. We would also like to thank CWJ Beenakker for his support.
MS was supported by the Netherlands Organization for
Scientific Research (NWO/OCW) and an ERC Synergy Grant. SW was supported by  STW Netherlands, an NWO VIDI Grant and an ERC Synergy Grant.
\bibliographystyle{unsrt}
\bibliography{ArticleQ}

\begin{thebibliography}{10}

\bibitem{olson2017quantum}
Jonathan Olson, Yudong Cao, Jonathan Romero, Peter Johnson, Pierre-Luc
  Dallaire-Demers, Nicolas Sawaya, Prineha Narang, Ian Kivlichan, Michael
  Wasielewski, and Al{\'a}n Aspuru-Guzik.
\newblock Quantum information and computation for chemistry.
\newblock {\em arXiv preprint arXiv:1706.05413}, 2017.

\bibitem{bravyi2002fermionic}
Sergey~B Bravyi and Alexei~Yu Kitaev.
\newblock Fermionic quantum computation.
\newblock {\em Annals of Physics}, 298(1):210--226, 2002.

\bibitem{aspuru2005simulated}
Al{\'a}n Aspuru-Guzik, Anthony~D Dutoi, Peter~J Love, and Martin Head-Gordon.
\newblock Simulated quantum computation of molecular energies.
\newblock {\em Science}, 309(5741):1704--1707, 2005.

\bibitem{whitfield2011simulation}
James~D Whitfield, Jacob Biamonte, and Al{\'a}n Aspuru-Guzik.
\newblock Simulation of electronic structure {H}amiltonians using quantum
  computers.
\newblock {\em Molecular Physics}, 109(5):735--750, 2011.

\bibitem{hastings2014improving}
Matthew~B Hastings, Dave Wecker, Bela Bauer, and Matthias Troyer.
\newblock Improving quantum algorithms for quantum chemistry.
\newblock {\em arXiv preprint arXiv:1403.1539}, 2014.

\bibitem{babbush2014adiabatic}
Ryan Babbush, Peter~J Love, and Al{\'a}n Aspuru-Guzik.
\newblock Adiabatic quantum simulation of quantum chemistry.
\newblock {\em Scientific reports}, 4, 2014.

\bibitem{lanyon2010towards}
Benjamin~P Lanyon, James~D Whitfield, Geoff~G Gillett, Michael~E Goggin,
  Marcelo~P Almeida, Ivan Kassal, Jacob~D Biamonte, Masoud Mohseni, Ben~J
  Powell, Marco Barbieri, et~al.
\newblock Towards quantum chemistry on a quantum computer.
\newblock {\em Nature chemistry}, 2(2):106--111, 2010.

\bibitem{du2010nmr}
Jiangfeng Du, Nanyang Xu, Xinhua Peng, Pengfei Wang, Sanfeng Wu, and Dawei Lu.
\newblock Nmr implementation of a molecular hydrogen quantum simulation with
  adiabatic state preparation.
\newblock {\em Physical review letters}, 104(3):030502, 2010.

\bibitem{peruzzo2014variational}
Alberto Peruzzo, Jarrod McClean, Peter Shadbolt, Man-Hong Yung, Xiao-Qi Zhou,
  Peter~J Love, Al{\'a}n Aspuru-Guzik, and Jeremy~L O’brien.
\newblock A variational eigenvalue solver on a photonic quantum processor.
\newblock {\em Nature communications}, 5, 2014.

\bibitem{wang2015quantum}
Ya~Wang, Florian Dolde, Jacob Biamonte, Ryan Babbush, Ville Bergholm, Sen Yang,
  Ingmar Jakobi, Philipp Neumann, Al{\'a}n Aspuru-Guzik, James~D Whitfield,
  et~al.
\newblock Quantum simulation of helium hydride cation in a solid-state spin
  register.
\newblock {\em ACS nano}, 9(8):7769--7774, 2015.

\bibitem{o2016scalable}
PJJ O’Malley, Ryan Babbush, ID~Kivlichan, Jonathan Romero, JR~McClean, Rami
  Barends, Julian Kelly, Pedram Roushan, Andrew Tranter, Nan Ding, et~al.
\newblock Scalable quantum simulation of molecular energies.
\newblock {\em Physical Review X}, 6(3):031007, 2016.

\bibitem{kandala2017hardware}
Abhinav Kandala, Antonio Mezzacapo, Kristan Temme, Maika Takita, Jerry~M Chow,
  and Jay~M Gambetta.
\newblock Hardware-efficient quantum optimizer for small molecules and quantum
  magnets.
\newblock {\em arXiv preprint arXiv:1704.05018}, 2017.

\bibitem{wecker2013can}
Dave Wecker, Bela Bauer, Bryan~K Clark, Matthew~B Hastings, and Matthias
  Troyer.
\newblock Can quantum chemistry be performed on a small quantum computer.
\newblock {\em arXiv preprint arXiv:1312.1695}, page~15, 2013.

\bibitem{poulin2014trotter}
David Poulin, Matthew~B Hastings, Dave Wecker, Nathan Wiebe, Andrew~C Doherty,
  and Matthias Troyer.
\newblock The {T}rotter step size required for accurate quantum simulation of
  quantum chemistry.
\newblock {\em arXiv preprint arXiv:1406.4920}, 2014.

\bibitem{babbush2015chemical}
Ryan Babbush, Jarrod McClean, Dave Wecker, Al{\'a}n Aspuru-Guzik, and Nathan
  Wiebe.
\newblock Chemical basis of {T}rotter-{S}uzuki errors in quantum chemistry
  simulation.
\newblock {\em Physical Review A}, 91(2):022311, 2015.

\bibitem{reiher2017elucidating}
Markus Reiher, Nathan Wiebe, Krysta~M Svore, Dave Wecker, and Matthias Troyer.
\newblock Elucidating reaction mechanisms on quantum computers.
\newblock {\em Proceedings of the National Academy of Sciences}, page
  201619152, 2017.

\bibitem{shor1999polynomial}
Peter~W Shor.
\newblock Polynomial-time algorithms for prime factorization and discrete
  logarithms on a quantum computer.
\newblock {\em SIAM review}, 41(2):303--332, 1999.

\bibitem{harrow2009quantum}
Aram~W Harrow, Avinatan Hassidim, and Seth Lloyd.
\newblock Quantum algorithm for linear systems of equations.
\newblock {\em Physical review letters}, 103(15):150502, 2009.

\bibitem{abrams1997simulation}
Daniel~S Abrams and Seth Lloyd.
\newblock Simulation of many-body {F}ermi systems on a universal quantum
  computer.
\newblock {\em Physical Review Letters}, 79(13):2586, 1997.

\bibitem{wigner1928uber}
Eugene~P Wigner and Pascual Jordan.
\newblock \"{U}ber das {P}aulische \"{A}quivalenzverbot.
\newblock {\em Z. Phys}, 47:631, 1928.

\bibitem{somma2002simulating}
Rolando Somma, Gerardo Ortiz, James~E Gubernatis, Emanuel Knill, and Raymond
  Laflamme.
\newblock Simulating physical phenomena by quantum networks.
\newblock {\em Physical Review A}, 65(4):042323, 2002.

\bibitem{somma2003quantum}
Rolando Somma, Gerardo Ortiz, Emanuel Knill, and James Gubernatis.
\newblock Quantum simulations of physics problems.
\newblock {\em International Journal of Quantum Information}, 1(02):189--206,
  2003.

\bibitem{verstraete2005mapping}
Frank Verstraete and J~Ignacio Cirac.
\newblock Mapping local {H}amiltonians of fermions to local {H}amiltonians of
  spins.
\newblock {\em Journal of Statistical Mechanics: Theory and Experiment},
  2005(09):P09012, 2005.

\bibitem{Note1}
We slightly abuse the nomenclature of quantum chemistry and molecular physics
  in merging spatial and spin quantum numbers into one index $j$, and use it as
  a label for what we call now the $j$-th orbital.

\bibitem{seeley2012bravyi}
Jacob~T Seeley, Martin~J Richard, and Peter~J Love.
\newblock The {B}ravyi-{K}itaev transformation for quantum computation of
  electronic structure.
\newblock {\em The Journal of chemical physics}, 137(22):224109, 2012.

\bibitem{tian2007constant}
Chao Tian, Vinay~A Vaishampayan, and NJA Sloane.
\newblock Constant weight codes: a geometric approach based on dissections.
\newblock {\em arXiv preprint arXiv:0706.1217}, 2007.

\bibitem{bravyi2016trading}
Sergey Bravyi, Graeme Smith, and John~A Smolin.
\newblock Trading classical and quantum computational resources.
\newblock {\em Physical Review X}, 6(2):021043, 2016.

\bibitem{moll2016optimizing}
Nikolaj Moll, Andreas Fuhrer, Peter Staar, and Ivano Tavernelli.
\newblock Optimizing qubit resources for quantum chemistry simulations in
  second quantization on a quantum computer.
\newblock {\em Bulletin of the American Physical Society}, 61, 2016.

\bibitem{bravyi2017tapering}
Sergey Bravyi, Jay~M Gambetta, Antonio Mezzacapo, and Kristan Temme.
\newblock Tapering off qubits to simulate fermionic {H}amiltonians.
\newblock {\em arXiv preprint arXiv:1701.08213}, 2017.

\bibitem{romero2016quantum}
Jonathan Romero, Jonathan Olson, and Alan Aspuru-Guzik.
\newblock Quantum autoencoders for efficient compression of quantum data.
\newblock {\em arXiv preprint arXiv:1612.02806}, 2016.

\bibitem{tranter2015bravyi}
Andrew Tranter, Sarah Sofia, Jake Seeley, Michael Kaicher, Jarrod McClean, Ryan
  Babbush, Peter~V Coveney, Florian Mintert, Frank Wilhelm, and Peter~J Love.
\newblock The {B}ravyi--{K}itaev transformation: {P}roperties and applications.
\newblock {\em International Journal of Quantum Chemistry}, 115(19):1431--1441,
  2015.

\bibitem{Note2}
`Unencoded state' is actually a slightly misleading term: when we say a state
  $\protect \boldsymbol {\lambda }\in \protect \mathbb {Z}_2^{\otimes N}$ is
  not encoded, we actually mean that it cannot be encoded and correctly
  decoded, so $\protect \boldsymbol {d}\left (\protect \boldsymbol {e}\left
  (\protect \boldsymbol {\lambda }\right )\right )\not =\protect \boldsymbol
  {\lambda }$.

\bibitem{ruckenstein1987mean}
Andrei~E Ruckenstein, Peter~J Hirschfeld, and J~Appel.
\newblock Mean-field theory of high-{T} c superconductivity: {T}he
  superexchange mechanism.
\newblock {\em Physical Review B}, 36(1):857, 1987.

\bibitem{Note3}
For better or worse we have used the binary representation of the orbital
  indexes. We could however employ any other counting method, i.e. any
  injective mapping that relates a binary vector representing a qubit basis
  state to an integer labeling an orbital.

\bibitem{barenco1995elementary}
Adriano Barenco, Charles~H Bennett, Richard Cleve, David~P DiVincenzo, Norman
  Margolus, Peter Shor, Tycho Sleator, John~A Smolin, and Harald Weinfurter.
\newblock Elementary gates for quantum computation.
\newblock {\em Physical review A}, 52(5):3457, 1995.

\bibitem{whitfield2016local}
James~D Whitfield, Vojt{\v{e}}ch Havl{\'\i}{\v{c}}ek, and Matthias Troyer.
\newblock Local spin operators for fermion simulations.
\newblock {\em Physical Review A}, 94(3):030301, 2016.

\bibitem{havlivcek2017operator}
Vojt{\v{e}}ch Havl{\'\i}{\v{c}}ek, Matthias Troyer, and James~D Whitfield.
\newblock Operator locality in the quantum simulation of fermionic models.
\newblock {\em Physical Review A}, 95(3):032332, 2017.

\bibitem{kitaev1995quantum}
A~Yu Kitaev.
\newblock Quantum measurements and the {A}belian stabilizer problem.
\newblock {\em arXiv preprint quant-ph/9511026}, 1995.

\bibitem{cleve1998quantum}
Richard Cleve, Artur Ekert, Chiara Macchiavello, and Michele Mosca.
\newblock Quantum algorithms revisited.
\newblock In {\em Proceedings of the Royal Society of London A: Mathematical,
  Physical and Engineering Sciences}, volume 454, pages 339--354. The Royal
  Society, 1998.

\bibitem{abrams1999quantum}
Daniel~S Abrams and Seth Lloyd.
\newblock Quantum algorithm providing exponential speed increase for finding
  eigenvalues and eigenvectors.
\newblock {\em Physical Review Letters}, 83(24):5162, 1999.

\bibitem{mcclean2016theory}
Jarrod~R McClean, Jonathan Romero, Ryan Babbush, and Al{\'a}n Aspuru-Guzik.
\newblock The theory of variational hybrid quantum-classical algorithms.
\newblock {\em New Journal of Physics}, 18(2):023023, 2016.

\bibitem{berry2015simulating}
Dominic~W Berry, Andrew~M Childs, Richard Cleve, Robin Kothari, and Rolando~D
  Somma.
\newblock Simulating {H}amiltonian dynamics with a truncated {T}aylor series.
\newblock {\em Physical review letters}, 114(9):090502, 2015.

\bibitem{berry2015hamiltonian}
Dominic~W Berry, Andrew~M Childs, and Robin Kothari.
\newblock Hamiltonian simulation with nearly optimal dependence on all
  parameters.
\newblock In {\em Foundations of Computer Science (FOCS), 2015 IEEE 56th Annual
  Symposium on}, pages 792--809. IEEE, 2015.

\bibitem{babbush2016exponentially}
Ryan Babbush, Dominic~W Berry, Ian~D Kivlichan, Annie~Y Wei, Peter~J Love, and
  Al{\'a}n Aspuru-Guzik.
\newblock Exponentially more precise quantum simulation of fermions in second
  quantization.
\newblock {\em New Journal of Physics}, 18(3):033032, 2016.

\bibitem{suzuki1990fractal}
Masuo Suzuki.
\newblock Fractal decomposition of exponential operators with applications to
  many-body theories and monte carlo simulations.
\newblock {\em Physics Letters A}, 146(6):319--323, 1990.

\bibitem{suzuki1991general}
Masuo Suzuki.
\newblock General theory of fractal path integrals with applications to
  many-body theories and statistical physics.
\newblock {\em Journal of Mathematical Physics}, 32(2):400--407, 1991.

\bibitem{nielsen2002quantum}
Michael~A Nielsen and Isaac Chuang.
\newblock Quantum computation and quantum information, 2002.

\end{thebibliography}
\onecolumngrid
\newpage
\appendix
\section{On quantum simulation}
\label{sec:hamilsim}

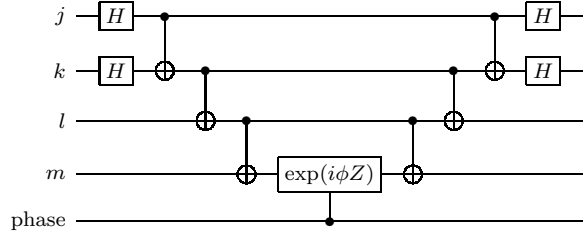
\begin{figure} \footnotesize
$\qquad$ \Qcircuit @C=1em @R=1.2em {
\lstick{j}	&\gate{H}	&\ctrl{1}	&\qw	&\qw	&\qw	&\qw	&\qw	&\ctrl{1}	&\gate{H}	& \qw\\
\lstick{k}	&\gate{H}	&\targ	&\ctrl{1}	&\qw	&\qw	&\qw	&\ctrl{1}	&\targ	&\gate{H}	& \qw\\
\lstick{l}	&\qw	&\qw	&\targ	&\ctrl{1}	&\qw	&\ctrl{1}	&\targ	&\qw	&\qw	&\qw\\
\lstick{m}	&\qw	&\qw	&\qw	&\targ	&\gate{\mathrm{exp}(i \phi Z)}	&\targ	&\qw	&\qw	&\qw	&\qw\\
\lstick{\mathrm{phase}}	&\qw	&\qw	&\qw	&\qw	&\ctrl{-1}	&\qw	&\qw	&\qw	&\qw & \qw											}
\caption{Implementing   $\mathrm{exp}\left( i \phi \, X_j \otimes X_k \otimes Z_l \otimes Z_m \right)$, conditional on qubit `phase'.  $\phi=\mathsf{s}\,\theta_{_{X_j \otimes X_k \otimes Z_l \otimes Z_m}}$ is a real rotation angle, where $\mathsf{s}$, is a time slice, and $\theta_{_{X_j \otimes X_k \otimes Z_l \otimes Z_m}}$ is the Hamiltonian weight of the string $X_j \otimes X_k \otimes Z_l \otimes Z_m $, as in  \eqref{eq:paulistringhamil}.
} \label{fig:gadget}

\end{figure}

 At this point, we discuss quantum simulation in the context of our transformations. Amongst other things, we describe the most simple algorithm for Hamiltonian simulation, and proceed by investigating feasibility issues with our transforms. Let us start by explaining how this work fits into the larger frame. \\
The transformations we have developed are going to be useful to trade quantum resources for quantum simulation of fermionic systems, independent from the concrete quantum algorithms chosen for simulation of the problem. For those problems from quantum chemistry and many-body physics we are  usually given a fermionic system and its Hamiltonian.  One is then to determine the system's ground state and ground state energy, sometimes parts of its spectrum. Where classical computation is infeasible, we simulate the system inside a quantum computer, on which the problem can be solved with existing algorithms.
 With either transform (see Table \ref{tab:othertable}), the fermionic system is therefore mapped to a system of $n$ qubits. With the operator transform, $H$  turns into $\mathsf{H}$, a sum of weighted $\mathcal{L}( ( \mathbb{C}^2)^{\otimes n} )$ gates, Pauli strings at best.
We then apply algorithms like quantum phase estimation \cite{kitaev1995quantum,cleve1998quantum,abrams1999quantum,aspuru2005simulated},  variational quantum eigensolvers  \cite{peruzzo2014variational,mcclean2016theory,o2016scalable,kandala2017hardware} , or adiabatic simulations \cite{babbush2014adiabatic}. All of those algorithms receive ansatz states as  inputs and  in some way prepare (eigen-) states, while also outputting their energy. The ground state is the state with the lowest energy, and can be then  manipulated as it is inside the quantum registers after the simulation.
 For the remainder of this Appendix, we discuss implications of the simulation algorithms onto our transforms. Thus we outline some principles, these algorithms rely on: algorithms might require us to simulate the  time evolution of our encoded system according to  $\mathsf{H}$. For that purpose, we need to know how to transform   the  time evolution operator $\mathrm{exp}\left( i \mathsf{H}\mathsf{t} \right)$, where $\mathsf{t}$ is a time step, into gate sequences. Maybe we even need to apply those evolution conditionally, means as an operation controlled on an auxiliary qubit (register). We thus need to embed $\mathsf{H}$  into an algorithm for Hamiltonian simulation.  \\
 Let us now be a bit more concrete, and select such an algorithm.
 Despite the wide range of theoretical proposals for Hamiltonian simulation algorithms \cite{berry2015simulating,berry2015hamiltonian,babbush2016exponentially}, only the perhaps simplest scheme appears to be experimentally feasible for digital quantum simulation at the moment. Note that it can only be applied to Hamiltonians that are a sum of Pauli strings with real weights
 \begin{align}
 \label{eq:paulistringhamil}
 \mathsf{H}\;=\;\sum_{\mathclap{\sigma \in \left\lbrace X,Y,Z,\mathbb{I} \right\rbrace^{\otimes n}}} \; \theta_\sigma \times\sigma \qquad \text{with all} \quad \theta_\sigma \in \mathbb{R} \, .
 \end{align}
The idea is to approximate $\mathrm{exp}\left( i \mathsf{H}\mathsf{t} \right)$, by sequences of the exponentiated Pauli strings $\mathrm{exp}\left( i \theta_\sigma \,\mathsf{s}\, \sigma \right)$, where $\mathsf{s}$ is a time slice of $\mathsf{t}$.  This method is commonly referred  to as Trotterization. The numbers, signs and values of the time slices $\mathsf{s}$, as well as the ordering of the exponentiated strings,  govern the error of the simulation -  strategies to minimize that error can be learned from the works of Suzuki \cite{suzuki1990fractal,suzuki1991general}. Note that we do not specify whether the Hamiltonian simulation is performed in an analog or digital fashion, however, not all strings $\sigma$ are feasible to be implemented in an analog fashion. The digital gadget for the exponentiation of Pauli strings, on the other hand,  is well known \cite{nielsen2002quantum}.  See Figure \ref{fig:gadget} for an example. We are therefore able to approximately perform a (conditional) simulated time evolution with $\mathsf{H}$ of the form \eqref{eq:paulistringhamil}. Using algorithms like variational eigensolvers, where we do not simulate the  time evolution but estimate the Hamiltonian expectation value by measuring its terms, we are in principle not tied to the structure of \eqref{eq:paulistringhamil}. However, it is more convenient.  \eqref{eq:paulistringhamil} gives us two constraints on how to transform \eqref{eq:generichamiltonian}. \\
The first constraint is that we need to decompose  every fermionic operator into Pauli strings, using \eqref{eq:prop6}. The total  number of Pauli strings resulting can be a problematically high when the underlying codes are highly non-linear. For  Trotterization that means a tremendous increase in length due to the abundance of sequenced Pauli string gadgets, many of them with very small rotation angles ($\phi$ in Figure \ref{fig:gadget}).\\
The second constraint seems trivial at first: in order to simulate a Hamiltonian, it has to be hermitian. More precisely, it has to be hermitian on the entire $(\mathbb{C}^2)^{\otimes n}$, so the coefficient $\theta_{\sigma}$ have to be real. We on the other hand might  not even  need the entire $(\mathbb{C}^2)^{\otimes n}$ to encode our physical system. Non-hermicities, meaning complex coefficients $\theta_\sigma$, can occur whenever one is careless with the remainder of the qubit space, when the code space is left or states are encoded in an ambiguous way.
 We here list a few pitfalls that can cause  non-hermitian terms to occur after the transform and discuss how to avoid them.
\begin{itemize}
\item Issues may be caused by  codes  that are not one-to-one. A one-to-one code ($\bbs{e}$, $\bbs{d}$) has the property: $\bbs{e}\left(\bbs{d}\left(\bbs{\omega}\right)\right)=\bbs{\omega}$ for all $\bbs{\omega} \in \mathbb{Z}_2^{\otimes n}$. Although we have excluded the one-to-one property from the definition of the codes (taing into account the next item), it assures the hermiticity of the transformed Hamiltonian.
\item  The encoded basis $\mathcal{B}$ has a size that is in between $2^n$ and $2^{n-1}$, so $n$ qubits provide too much Hilbert space by default. However, we can always add a state to the basis that is mapped to zero by all terms $\hat{h}_{\bbs{ab}}$.  This state, represented by $\bbs{\nu}$, can have several partners on the code space $\bbs{\omega}$, for which $\bbs{d}\left( \bbs{\omega} \right)=\bbs{\nu}$ (i.e. not be mapped one-to-one). For example for particle-number conserving Hamiltonians, we can balance these dimensional mismatches using the  vacuum state in such a way, since  $c^{\dagger}_ic^{\;}_j\ket{\Theta}=0$.
\item  We encounter this problem when using a code with a Hamiltonian, that is not feasible with it. The segment codes for instance are feasible only for certain adjusted particle-number-conserving Hamiltonians, as we shall see in  Appendix \ref{sec:ments}.
\end{itemize}

\section{General operator mappings}
\label{sec:opmap}
The goal of this Appendix is to verify that the fermionic mode is accurately represented by our qubit system. This is divided into three steps: step one is to analyze the action of Hamiltonian terms on the fermionic basis. In the second step, we verify parity and projector parts of \eqref{eq:admiral} to work like the original operators in step one, disregarding the occupational update for a moment. Conditions for this state update are subsequently derived. The update operator \eqref{eq:updateop} is shown to fulfill these conditions in the third step, thus concluding the proof.

\subsection{Hamiltonian dynamics}
In order to verify that  the gate sequences \eqref{eq:admiral} are  mimicking the Hamiltonian dynamics adequately, we verify that the resulting terms have the same effect on the Hamiltonian basis.
 This is done on the level of second quantization with respect to the notation \eqref{eq:btov}: no transition into a qubit system is made. This step serves the sole purpose to  quantify the effect of the Hamiltonian terms on the states.
To that end, we begin by  studying the effect of a singular fermionic  operator $c^{(\dagger)}_j$ on a pure state, before considering an entire  term $\hat{h}_{\bbs{ab}}$ on a state in $\mathcal{B}$. As a preliminary, we note that \eqref{eq:stateaction1}-\eqref{eq:stateaction4} follow directly from \eqref{eq:antiraw}, when considering that
\begin{align}
c^{\;}_j c^{\;}_j = c^{\dagger}_j c^{\dagger}_j = c^{\;}_j \ket{\Theta} = 0 \, .
\end{align}
  The relations \eqref{eq:stateaction1}-\eqref{eq:stateaction4} indicate how singular operators act on pure states in general. We now become more specific and apply these rules to a state $(\prod_{i} (c_i^\dagger)^{\nu_i}) \ket{\Theta}$, that is not necessarily in $\mathcal{B}$, but is described by an occupation vector  $\bbs{\nu} \in \mathbb{Z}_2^{\otimes N}$. The effect of an annihilation operator on such a state is considered first:
 \begin{align}\label{eq:jwstep0}
c_j^{\,} \left[\prod_{i=1}^{N} \left(c_i^\dagger\right)^{\nu_i}\right] \ket{\Theta} = &\left[\prod_{i<j}\left(-c_i^\dagger\right)^{\nu_i}\right]\;\; c_j^{\,}\left(c_j^\dagger\right)^{\nu_j}\; \; \left[\prod_{k>j}\left(c_k^\dagger\right)^{\nu_k} \right] \ket{\Theta}\; \\ \label{eq:jwstep1}
 =  &\left[\prod_{i<j}\left(-c_i^\dagger\right)^{\nu_i }\right]\;\; \frac{1}{2}\left[1-\left(-1\right)^{\nu_j} \right]\; \; \left[\prod_{k>j}\left(c_k^\dagger\right)^{\nu_k} \right] \ket{\Theta}\; \\ \label{eq:jwstep2}
 =  &\left[\prod_{i<j}\left(-1\right)^{\nu_i }\right]\;\; \frac{1}{2}\left[1-\left(-1\right)^{\nu_j} \right]\; \; \left[\prod_{k=1}^{N}\left(c_k^\dagger\right)^{\nu_k + \delta_{jk} \moto}\right] \ket{\Theta}\;
\end{align}
  A short explanation on what has happened:
in \eqref{eq:jwstep0}, $c_j$ has anticommuted with all creation operator $c_i^\dagger$ that have  indexes $i<j$.  Depending on the component $\nu_j$, a creation operator $c_j^\dagger$ might now be to the right of the annihilator $c_j$. If the creation operator is not encountered, we may continue the anticommutations of $c_j$ until it  meets the vacuum and annihilates the state by $c_j\ket{\Theta}=0$. Using the anticommutation relations \eqref{eq:antiraw}, we therefore replace $c_j^{\,}(c_j^\dagger)^{\nu_j}$ with $\frac{1}{2}\left[1-(-1)^{\nu_j} \right]$ when going from \eqref{eq:jwstep0} to \eqref{eq:jwstep1}. Finally, the terms are rearranged in \eqref{eq:jwstep2}: conditional sign changes of the anticommutations are factored out of the new state with an occupation that is now described by the binary vector $( \bbs{\nu}+\bbs{u_j} \moto)$ rather than  $\bbs{\nu}$. When considering to apply a creation operator $c_j^\dagger$ on the former state, the result is similar. Alone at step \eqref{eq:jwstep1}, we have to replace $c^\dagger_j (c^\dagger_j )^{\nu_j}$ by $\frac{1}{2}\left[1+\left(-1\right)^{\nu_j} \right]$ instead, as now the case of appearance of the creation operator leads to annihilation: $c^\dagger_jc^\dagger_j=0$. We thus find
\begin{align}
\label{eq:jwstep3}
c_j^{\dagger} \left[\prod_{i=1}^{N} \left(c_i^\dagger\right)^{\nu_j}\right] \ket{\Theta} = \left[\prod_{i<j}\left(-1\right)^{\nu_i}\right]\;\; \frac{1}{2}\left[1+\left(-1\right)^{\nu_j} \right]\; \; \left[\prod_{k=1}^{N}\left(c_k^\dagger\right)^{\nu_k + \delta_{jk} \moto}\right] \ket{\Theta}\; .
\end{align}
We now turn our attention to the actual goal,  effect of a Hamiltonian term from \eqref{eq:generichamiltonian} on a state in $\mathcal{B}$ (this means its occupation vector $\bbs{\nu}$ is in $\mathcal{V}$).
 We therefore consider a generic operator sequence $\prod_{i=1}^l (c^{\dagger}_{a_i} )^{b_i}(c^{\,}_{a_i})^{1+b_i \moto}$, parametrized by some $N$-ary vector $\bbs{a} \in [N]^{\otimes l}$ and a binary vector $\bbs{b}\in \mathbb{Z}_2^{\otimes l}$, for some length $l$.
With \eqref{eq:jwstep2} and \eqref{eq:jwstep3}, we now have the means to consider the effect such a sequence of annihilation and creation operators.  The two relations will be repeatedly utilized in an inductive procedure, as every single operator $(c^{\dagger}_{a_i} )^{b_i}(c^{\,}_{a_i})^{1+b_i \moto}$ of  $\prod_{i=1}^l (c^{\dagger}_{a_i} )^{b_i}(c^{\,}_{a_i})^{1+b_i \moto}$ will act on a basis state, one after another. The state's occupation is updated after every such operation. For convenience, we define:
\begin{align}
 \bbs{\nu^{(i)}}&\left. \in \mathbb{Z}_2^{\otimes N}\quad  \right| \;  i \in  \lbrace 0, \,\dots ,\, l \rbrace  \label{eq:iterate-1}\\
\bbs{\nu^{(l)}}\,&=\,\bbs{\nu} \in \mathcal{V} \label{eq:iterate0}\\
\bbs{\nu^{(i- 1)}}\,&=\, \bbs{\nu^{(i)}} + \bbs{u_{a_{i} }}\;\moto \label{eq:iterate} \, .
\end{align}
Now, the procedure starts:
 \begin{align}
&\left[\,\prod_{i=1}^l \left(c^{\dagger}_{a_i} \right)^{b_i}\left(c^{\,}_{a_i}\right)^{1+b_i \moto}\right] \;  \left[\,\prod_{k=1}^{N} \left( c_k^\dagger \right)^{\nu_k}\right] \ket{\Theta} \\
=&\left[\,\prod_{i=1}^{l-1}\left(c^{\dagger}_{a_i} \right)^{b_i}\left(c^{\,}_{a_i}\right)^{1+b_i \moto}\right] \; \frac{1}{2}\left[1-\left(-1\right)^{b_l}(-1)^{\nu_{a_l}} \right]\left(-1\right)^{\sum_{j<a_l}\nu_j} \left[\, \prod_{k=1}^{N} \left( c_k^\dagger \right)^{\nu_k + \delta_{a_l k} \moto } \right]\ket{\Theta} \label{eq:induction0}\\
=&\left[\frac{1}{2}\left[1-\left(-1\right)^{b_l}(-1)^{\nu_{a_l}^{(l)}} \right]\left(-1\right)^{\sum_{j<a_l}\nu^{(l)}_j}\right] \;\left[\,\prod_{i=1}^{l-1}\left(c^{\dagger}_{a_i} \right)^{b_i}\left(c^{\,}_{a_i}\right)^{1+b_i \moto}\right] \; \left[\, \prod_{k=1}^{N} \left( c_k^\dagger \right)^{\nu^{(l-1)}_k } \right]\ket{\Theta}  \label{eq:induction1}\\
=&\left[\,\prod_{i=1}^{l} \right. \underbrace{\frac{1}{2}\left[1-\left(-1\right)^{b_{i}}(-1)^{\nu^{(i)}_{a_{i}}} \right]}_{\text{projector eigenvalues}}\underbrace{\left(-1\right)^{\sum_{j<a_{i}}\nu^{(i)}_j}}_{\text{parity signs}} \left. \vphantom{\,\prod_{i=1}^{l}} \right] \label{eq:induction2}\;\underbrace{\left[\,\prod_{k=1}^{N} \left( c_k^\dagger \right)^{\nu^{(0)}_k}\right] \ket{\Theta}}_{\text{updated state}}
\end{align}
We again explain what has happened:
first, the rightmost operator, which is either $c^{\;}_{a_l}$ or $c^{\dagger}_{a_l}$ depending on the parameter  $b_l$, acts on the state according to either \eqref{eq:jwstep2} or \eqref{eq:jwstep3}. We therefore combine the two relations for the absorption of this operator  $(c^{\dagger}_{a_l})^{b_l}(c^{\,}_{a_l})^{1+b_l \moto}$ in \eqref{eq:induction0}. In the same fashion, all the remaining operators of the sequence are one-after-another absorbed into the state.
 The new state is described by the vector $\bbs{\nu^{(l-1)}}$ after the update. And the cycle begins anew with $(c^{\dagger}_{a_{l-1}} )^{b_{l-1}}(c^{\,}_{a_{l-1}})^{1+b_{l-1} \moto}$. From \eqref{eq:induction1} on, we use the notations \eqref{eq:iterate-1}-\eqref{eq:iterate} to describe partially updated occupations.  By the end of this iteration, the occupation of the state is changed to  $~\bbs{\nu^{(0)}}~=~\bbs{\nu}+\bbs{q}\moto\;$,  with the total change  $\;\bbs{q}=\sum_i\bbs{u_{a_i}}\moto$.   Also, the coefficients of \eqref{eq:induction2} take into account sign changes from  anticommutations (``parity signs" in \eqref{eq:induction2}) and the eigenvalues of the applied projections. In its entirety,
\eqref{eq:induction2} denotes the resulting state, and is the main ingredient for the next step.
\subsection{Parity operators and projectors}
We are given the operator transform \eqref{eq:admiral} and the state transform \eqref{eq:statemap}. We want to show the that the fermion system is adequately simulated, which means to show that the effect \eqref{eq:induction2} is replicated by \eqref{eq:admiral} acting on $\ket{\bbs{e}(\bbs{\nu})}$.  This is the goal of the next two steps. We start by evaluating the application of \eqref{eq:admiral} on that state, up to the update operator $\mathcal{U}^{\, a}$. This means that the operators applied implement two things only: the parity signs of \eqref{eq:induction2}, and the projection onto the correct occupational state. Note that these parity operators and projectors are applied before the update operator in \eqref{eq:admiral}:
\begin{align}
\underbrace{\vphantom{\left( \prod_{v=1}^{l-1} \right)}\mathcal{U}^{\,\bbs{a}}}_{\text{update operator}} \underbrace{\left( \prod_{v=1}^{l-1}\,\prod_{w=v+1}^{l}\left(-1 \right)^{\theta_{a_v a_w}} \right)}_{\text{parity operators}}  \prod_{x=1}^{l}\underbrace{ \frac{1}{2}\left(\mathbb{I}- \left[\prod_{y=x+1}^{l}(-1)^{\delta_{a_x a_y}}\right](-1)^{b_x}\;\mathfrak{X}\left[d_{a_x}\right]\right)}_{\text{projectors}} \underbrace{\vphantom{\left( \prod_{v=1}^{l-1} \right)}\mathfrak{X}\left[ \,p_{a_x} \right]}_{\text{parity operators}} \, .
\end{align}
 We now commence our evaluation:
\begin{align}
&  \mathcal{U}^{\,\bbs{a}} \left[\left( \prod_{v=1}^{l-1}\,\prod_{w=v+1}^{l}\left(-1 \right)^{\theta_{a_v a_w}} \right)  \prod_{x=1}^{l} \frac{1}{2}\left(\mathbb{I}- \left[\prod_{y=x+1}^{l}(-1)^{\delta_{a_x a_y}}\right](-1)^{b_x}\;\mathfrak{X}\left[d_{a_x}\right]\right) \mathfrak{X}\left[ \,p_{a_x} \right] \right]\quad \ket{\bbs{e}\left(\bbs{\nu} \right)}  \\
=&\label{eq:verify1}\;\mathcal{U}^{\,\bbs{a}} \left[\left( \prod_{v=1}^{l-1}\,\prod_{w=v+1}^{l}\left(-1 \right)^{\theta_{a_v a_w}} \right)  \prod_{x=1}^{l} \frac{1}{2}\left(1- \left[\prod_{y=x+1}^{l}(-1)^{\delta_{a_x a_y}}\right](-1)^{b_x}\;{\left( -1\right)^{d_{a_x}\left( \bbs{e}\left(\bbs{\nu}\right)\right)}}\right){\left( -1\right)^{ p_{a_x}\left( \bbs{e}\left(\bbs{\nu}\right)\right)}}\right]\quad \ket{\bbs{e}\left(\bbs{\nu} \right)} \\
=&\label{eq:verify2}\;\mathcal{U}^{\,\bbs{a}} \left[\left( \prod_{v=1}^{l-1}\,\prod_{w=v+1}^{l}\left(-1 \right)^{\theta_{a_v a_w}} \right)  \prod_{x=1}^{l} \frac{1}{2}\left(1- \left[\prod_{y=x+1}^{l}(-1)^{\delta_{a_x a_y}}\right](-1)^{b_x}\;{\left(-1\right)^{\nu_{a_x}}}\right){\left( -1\right)^{\sum_{j<a_x} \nu_j}}\right]\quad \ket{\bbs{e}\left(\bbs{\nu} \right)}\\
=&\label{eq:verify3}\; \mathcal{U}^{\,\bbs{a}} \; \left[ \prod_{x=1}^{l} \frac{1}{2}\left(1- (-1)^{b_x}\;{\left(-1\right)^{\nu_{a_x}+\sum_{y=x+1}^{l}\delta_{a_x a_y}}}\right)\left( -1\right)^{\sum_{j<a_x} \nu_j+\sum_{y=x+1}^{l}\theta_{a_xa_y}}\right]\quad \ket{\bbs{e}\left(\bbs{\nu} \right)}\\
=&\label{eq:verify4}\; \phantom{\mathcal{U}^{\,\bbs{a}} \; }\left[\prod_{x=1}^{l} \frac{1}{2}\left(1- (-1)^{b_x}\;{\left(-1\right)^{\nu_{a_x}^{(x)}}}\right)\left( -1\right)^{\sum_{j<a_x} \nu_j^{(x)}} \right]\quad  \mathcal{U}^{\,\bbs{a}} \; \ket{\bbs{e}\left(\bbs{\nu} \right)} \, .
\end{align} Let us describe what has happened: in \eqref{eq:verify1}, the extraction property \eqref{eq:prop1} is used, and we arrive at \eqref{eq:verify2} after using the property $\bbs{d}\left(\bbs{e}\left(\bbs{\nu}\right) \right)=\bbs{\nu}$ and the definition of the parity function. From there we go to \eqref{eq:verify3} when we merge the two products and perform rearrangements that make it easy to cast all delta and theta functions into the components of the partially updated occupations  $\bbs{\nu^{(i)}}$, \eqref{eq:verify4}. \\
Comparing \eqref{eq:verify4} to \eqref{eq:induction2}, we notice to have successfully mimicked the same sign changes and and projections, as the coefficients in both relations match. Now it is only left to show that the state update is executed correctly. Naively, one would think that we would need to show that
\begin{align}
\label{eq:sooonaive}
\mathcal{U}^{\,\bbs{a}} \; \ket{\bbs{e}\left(\bbs{\nu} \right)} \hat{=} \left[\,\prod_{k=1}^{N} \left( c_k^\dagger \right)^{\nu^{(0)}_k}\right] \ket{\Theta} \, ,
\end{align}
but this is  too strong a statement.   It is  in fact sufficient to demand
\begin{align}
\label{eq:demand}
\mathcal{U}^{\, \bbs{a}}\ket{\bbs{e}\left( \bbs{\nu} \right)}\; = \; \ket{\bbs{e}\left( \bbs{\nu^{(0)}} \right)}\;  = \;\ket{\bbs{e}\left( \bbs{\nu}+\bbs{q} \moto\right)} \, .
\end{align}
 For $\bbs{\nu^{(0)}} \in \mathcal{V}$, \eqref{eq:sooonaive} and \eqref{eq:demand} is equivalent. However, it might be the case that $\bbs{\nu^{(0)}} \notin \mathcal{V}$, so $\bbs{\nu^{(0)}}$ is not encoded. This mean that \eqref{eq:sooonaive} is not fulfilled, since $\bbs{d}(\bbs{e}(\bbs{\nu^{(0)}}))\neq \bbs{\nu^{(0)}}$. It is however not necessary to include $\bbs{\nu^{(0)}}$ in the encoding, as for $\bbs{\nu^{(0)}} \notin \mathcal{V}$, the state will vanish anyways:  we know from $\hat{h}_{\bbs{ab}}: \mathrm{span}(\mathcal{B})\to \mathrm{span}(\mathcal{B})$, that in this case $\hat{h}_{\bbs{ab}}$ must act destructively on that basis state, $\hat{h}_{\bbs{ab}}\,(\prod_{k} ( c_k^\dagger)^{\nu_k}) \ket{\Theta}=0$. This detail is implemented by the projector part of the transformed sequence \eqref{eq:admiral}. These projectors are, as we have just shown, working faithfully like \eqref{eq:induction2}, for the transformed sequence acting on  every $\ket{\bbs{\nu}}$ with $\bbs{\nu} \in \mathcal{V}$. Hence \eqref{eq:demand} is a sufficient condition for the updated state.
  The proof is completed once we have verified that \eqref{eq:demand} is satisfied with the update operator defined as in \eqref{eq:updateop}. This is done during the next step.
\subsection{Update operator}
The missing piece of the proof is to  check  that  \eqref{eq:updateop} and \eqref{eq:linenc} fulfill the condition \eqref{eq:demand}.
We start by verifying the condition \eqref{eq:demand} for \eqref{eq:linenc}, which we have presented as special case of \eqref{eq:updateop} for linear encoding functions: $\bbs{e}\left( \bbs{\nu}+\bbs{\nu^\prime}\moto \right)= \bbs{e}\left( \bbs{\nu}\right)+\bbs{e}\left( \bbs{\nu^\prime}\right) \moto$. Using that property, one can in fact derive \eqref{eq:linenc} from \eqref{eq:updateop} directly. We now apply \eqref{eq:linenc} to $\ket{\bbs{e}( \bbs{\nu })}$, but firstly we note that
\begin{align}
\label{eq:imprint}
X_j \ket{\bbs{\omega}} = \ket{\bbs{\omega}+\bbs{u_j}\moto} \, ,
\end{align}
where $\bbs{u_j}$ is  the $j$-th unit vector of  $\mathbb{Z}_2^{\otimes n}$. Using \eqref{eq:imprint} and the linearity of $\bbs{e}$, we find:
\begin{align}
\label{eq:lineup}
\mathcal{U}^{\, \bbs{a}} \ket{\bbs{e}\left(\bbs{\nu} \right)} \;&=\;\left[\bigotimes_{i=1}^{n} \left(X_i\right)^{\sum_{j}A_{ij} q_j \moto} \right]\ket{\bbs{e}(\bbs{\nu})} \; \\ &= \;\left[\bigotimes_{i=1}^{n} \left(X_i\right)^{\bbs{e}(\bbs{q})} \right]\ket{\bbs{e}(\bbs{\nu})} \\ &=\; \ket{\bbs{e}(\bbs{\nu})+\bbs{e}(\bbs{q})\moto} \;\\ &=\; \ket{\bbs{e}(\bbs{\nu}+\bbs{q}\moto)} \, ,
\end{align}
which shows \eqref{eq:demand} for linear encodings. \\We now turn our attention to general encodings and prove the same expression for update operators as defined in \eqref{eq:updateop}:
\begin{align}
\label{eq:demand1}
\mathcal{U}^{\, \bbs{a}} \ket{\bbs{e}\left(\bbs{\nu} \right)} \;= \;& \left(\sum_{\bbs{t}\, \in\, \mathbb{Z}_2^{\otimes n}}\left[\bigotimes_{i=1}^{n} \left(X_i\right)^{t_i}\right] \prod_{j=1}^{n} \frac{1}{2}\left( \mathbb{I}+(-1)^{t_j}\;\mathfrak{X}\left[ \varepsilon^{\, \bbs{q}}_j \right]\right) \right) \;\ket{\bbs{e}\left( \bbs{\nu } \right)} \\
=\;& \label{eq:demand2}\left(\sum_{\bbs{t}\, \in\, \mathbb{Z}_2^{\otimes n}}\left[\bigotimes_{i=1}^{n} \left(X_i\right)^{t_i}\right] \right.\prod_{j=1}^{n} \underbrace{\frac{1}{2}\left( 1+(-1)^{t_j+\varepsilon^{\, \bbs{q}}_j\left(  \bbs{e}\left(\bbs{\nu} \right)\right) }\right)}_{\qquad\delta_{t^{\;}_j\varepsilon_j^{\bbs{q}}(\bbs{e}(\bbs{\nu}))}}\left. \vphantom{\sum_{\bbs{t}\, \in\, \mathbb{Z}_2^{\otimes n}}}\right) \;\ket{\bbs{e}\left( \bbs{\nu }  \right)} \\
=\;& \label{eq:demand3}\left(\bigotimes_{i=1}^{n} \left(X_i\right)^{\varepsilon^{\, \bbs{q}}_i\left(  \bbs{e}\left(\bbs{\nu} \right)\right)}\right) \; \ket{\bbs{e}\left( \bbs{\nu } \right)}\\
=\;&  \label{eq:demand4}\ket{\bbs{e}\left( \bbs{\nu } \right)+ \bbs{\varepsilon}^{\, \bbs{q}}\left(  \bbs{e}\left(\bbs{\nu} \right)\right) \moto} \\ =\;& \ket{\bbs{e}\left(\bbs{\nu} \right)+\bbs{e}\left(\bbs{\nu} \right)+\bbs{e}\left( \bbs{d} \left( \bbs{e} \left( \bbs{\nu} \right) \right) + \bbs{q} \moto \right) \moto} \\
=\;&\ket{\bbs{e}\left( \bbs{\nu}+\bbs{q}\moto\right)} \, ,
\end{align}
which completes the proof.
We swiftly recap what has happened: in \eqref{eq:demand1}, we have plugged the  definition of\eqref{eq:updateop}  into the left-hand side of \eqref{eq:demand}. In between this equation and \eqref{eq:demand2}, we have evaluated the expectation values of the extracted operators $\mathfrak{X}[\varepsilon_j^{\bbs{q}}]$. From that line to the next, the $\mathbb{Z}_2^{\otimes n}$-sum is collapsed over the condition $\bbs{t}=\bbs{\varepsilon}^{\, \bbs{q}}(  \bbs{e}(\bbs{\nu} ))$. We go from \eqref{eq:demand3} to \eqref{eq:demand4} by applying \eqref{eq:imprint}.  Once we insert  the definition \eqref{eq:varepsilon} into \eqref{eq:demand4}, it becomes obvious that the condition \eqref{eq:demand} is fulfilled. Thus, the entire operator transform is now proven.
 \section{Transforming particle-number conserving Hamiltonians }
 \label{sec:altmaps}
  In this Appendix, we examine the richest symmetry to exploit for qubit savings: particle conservation. We begin by introducing the most relevant class of Hamiltonians that exhibit this symmetry, but  ultimately the main goal of this Appendix is to simplify the operator transform for all such Hamiltonians.  Motivated by the compartmentalized recipes of the conventional mappings, \eqref{eq:line}, we suggest alternatives to the transform \eqref{eq:admiral}, that do not depend on the sequence length $l$.   \\
  Let us start by noting how easy it is to state that a Hamiltonian the total number of particles:  a Hamiltonian like \eqref{eq:generichamiltonian}, conserves the total number of particles when every term $\hat{h}_{\bbs{ab}}$ has as many creation operators as it has annihilation operators. The lengths $l$,  implicit in the sequences $\hat{h}_{\bbs{ab}}$ that occur in the  Hamiltonian, are thereby determined by the field theory or model, that underlies the problem. The coefficients $h_{\bbs{ab}}$, on the other hand, are determined by the set of  basis functions used. For first-principle problems in quantum chemistry and solid state physics, we usually encounter particle-number-conserving Hamiltonians with terms of weight that is at most $l=4$:
  \begin{align}
\label{eq:quchem}
H=\sum_{ij} t_{ij}\;  c_i^\dagger c_j^{\;} + \sum_{ijkl} U_{ijkl} \; c_i^\dagger c_j^{\dagger}c_k^{\;} c_l^{\;} \, ,
\end{align}
where $U_{ijkl}$, $t_{ij}$ are complex coefficients of the interaction and single particle terms, respectively. In the notation of \eqref{eq:generichamiltonian}, these coefficients correspond to $h_{(i,j,k,l)(1,1,0,0)}$ and $h_{(i,j)(1,0)}$. The $(l=4)$ interaction terms usually originate from either magnetism and/or the Coulomb interaction. Even for these $(l=4)$-terms, the operator transform \eqref{eq:admiral} is quite bulky, and we in general would like to have a transform that is independent of $l$. Before we begin to discuss such transform recipes however, we need to set up some preliminaries. First of all, we need to find a suitable code ($\bbs{e}$, $\bbs{d}$), as discussed in the main part. Ideally, we would encode only the Hilbert space with the correct number of particles, $M$, but  Hilbert spaces of other particle numbers can also be included. Assuming that the Hamiltonian visits every state with the same particle number, we must encode entire Hilbert spaces $\mathcal{H}_N^{m}$ only.   Secondly, we need to reorder the fermionic operators inside the Hamiltonian terms $\hat{h}_{\bbs{ab}}$. The reason for this is, that our goal can only be achieved by finding recipes for smaller sequences of constant length. In order to transform the Hamiltonian terms then, we need to invoke the anticommutation relations \eqref{eq:antiraw} to introduce an order in $\hat{h}_{\bbs{ab}}$, such that these small sequences appear as consecutive, distinct blocks. As we shall see, these blocks will have the shape $c_i^{\dagger}c_j^{\;}$. So every $\hat{h}_{\bbs{ab}}$ needs to be reordered, such that every even operator  is a creation operator, and every odd operator an annihilator. For the $(l=4)$-terms in \eqref{eq:quchem}, this reordering means
  $c_i^{\dagger}c_j^{\dagger} c_k^{\;}c_l^{\;}\;\to\;c_i^{\dagger}c_l^{\;}c_j^{\dagger}c_k^{\;}- \delta_{jl}\;c_i^{\dagger}c_k^{\;}$.\\ Let us quickly sketch the idea behind that reordering and introduce some nomenclature: instead of considering Hamiltonian terms, we realize that also the terms $c_i^{\dagger}c_j^{\;}$ also conserve the particle number: $\mathcal{H}_N^m\to \mathcal{H}_N^m$. Let us act with $c_i^{\dagger}c_j^{\;}$ on an encoded state. We consider a  state that is not annihilated by  $c_i^{\dagger}c_j^{\;}$.  Its particle number  is reduced by one through $c_j$, but then immediately restored by $c_i^{\dagger}$. In fact,  for a general sequence of that arrangement,  every even operator  restores the particle number in this way and every odd reduces it. We therefore call the  subspace, in which we find the state after an even (odd) number of operators, the even (odd) subspace.  Since all  $l$ must be even for the Hamiltonian to have particle conservation symmetry, the even subspace is the one encoded. The odd subspace, on the other hand, has one particle less, so it is $\mathcal{H}_N^{(M-1)}$, if the even one is $\mathcal{H}_N^{M}$.

\subsection{Encoding the two spaces separately }
\label{subsec:separate}
In this ordering, one can find a recipe for a singular  creation or annihilation operator. The strategy is to consider a second code for the odd subspace. As before ($\bbs{e}$, $\bbs{d}$) denotes the code for the even subspace, and now ($\bbs{e^\prime}$, $\bbs{d^\prime}$) is encoding the odd subspace.  The idea is that after an odd operator (which in this ordering is an annihilation operator), the state is updated into the odd subspace. With every even operator (which is a creation operator), the state is updated from the odd subspace back into the even one.  We find:
\begin{align}
\label{eq:single1}
c^\dagger_j \quad &\hat{=} \quad \frac{1}{2}\; \bar{\mathcal{U}}^{\,(j)}\; \left( \mathbb{I}+\mathfrak{X}\left[ d_j\right] \right) \; \mathfrak{X}\left[ p_j\right]\, ,  \\ \label{eq:single2} c^{\;}_j \quad &\hat{=} \quad \frac{1}{2}\; \mathcal{U}^{\,(j)}\; \left( \mathbb{I}-\mathfrak{X}\left[ d^\prime_j\right] \right)  \mathfrak{X}\left[ p^\prime_j\right] \, .
\end{align}
In \eqref{eq:single2},  $\mathcal{U}^{\,(j)}$ is defined as in \eqref{eq:updateop}, but its counterpart from \eqref{eq:single1} is defined by
\begin{align}
\bar{\mathcal{U}}^{\,(j)}=
 \sum_{\bbs{t}\, \in\, \mathbb{Z}_2^{\otimes n}}\left[\bigotimes_{i=1}^{n} \left(X_i\right)^{t_i}\right] \prod_{i=1}^{n} \frac{1}{2}\left( \mathbb{I}+(-1)^{t_i}\;\mathfrak{X}\left[ \varepsilon^{ \prime \; \bbs{u_j}}_k \right]\right) \, ,
\end{align}
with the  primed functions $\bbs{\varepsilon^{\prime\; q}}$, $\bbs{p^\prime}$ defined like \eqref{eq:varepsilon} and \eqref{eq:p}, but with ($\bbs{e^\prime}$,~$\bbs{d^\prime}$) in place of  ($\bbs{e}$,~$\bbs{d}$). \\
This method relies on $n$ qubits being feasible to simulate the odd subspace in. That is, however, not always the case. The  basis set  of $\mathcal{H}_N^{M-1}$ is in general larger than $\mathcal{H}_N^M$, when $M>N/2$. In this way, the odd subspace can also be larger and even be infeasible to simulate with just $n$ qubits.  As a solution, one changes the ordering into odd operators being creation operators, and even ones being annihilators, like    $c_k^{\;}c_i^{\dagger}\,c_l^{\;}c_j^{\dagger}$. This causes the odd subspace to become $\mathcal{H}_N^{(M+1)}$, which has a smaller basis set than $\mathcal{H}_N^{M}$. For that case ($\bbs{e}$,~$\bbs{d}$) become the code for the odd subspace, and ($\bbs{e^\prime}$,~$\bbs{d^\prime}$) will be associated to the even subspace in \eqref{eq:single1} and \eqref{eq:single2}.\\ The obvious disadvantage is that two codes have to be employed at once. However,  the \party  for instance (Section\ref{subsubsec:party} in the main part),  comes in two different flavors already, which can be used as codes for even and odd subspaces, respectively.
\subsection{Encoding the building blocks} The building blocks $c_i^\dagger c_j^{\;}$ are guaranteed to conserve the particle number, so the even subspace is conserved. As a consequence, one may consider the possibility to transform the operators as the pairs we have rearranged them into.  In this way, we still have a certain compartmentalization of \eqref{eq:admiral}.
 Two special cases are to be taken into account: when $i>j$, an additional minus sign has to be added, as compared to the $i<j$ case.  Also, when $i=j$, all parity operators cancel and the projectors coincide.   We find:
\begin{align}
\label{eq:trafo2}
c_i^\dagger c_j \; &\hat{=} \; \begin{cases}  \frac{1}{4} \; (-1)^{\theta_{ij}} \; \mathcal{U}^{(i,j)}\; \mathfrak{X}\left[p_i\right]\; \mathfrak{X}\left[p_j\right]\;\left(\mathbb{I}+\mathfrak{X}\left[d_i\right]\right)\left(\mathbb{I}-\mathfrak{X}\left[d_j\right]\right)  & i\neq j \\ \, \\
\frac{1}{2}\left(1-\mathfrak{X}\left[d_j\right]\right)& i=j \, ,
\end{cases}
\end{align}
with $ \mathcal{U}^{(i,j)}$ being the  $l=2$ version of \eqref{eq:updateop}, and $\bbs{p}$ and $\bbs{\varepsilon^{\, q}}$  defined as usual  by \eqref{eq:p} and \eqref{eq:varepsilon}.
\section{Multi-weight binary addressing codes based on dissections}
\label{sec:binary}
With binary addressing codes, that is codes that are similar to the one presented in Section \ref{subsubsec:binadressing} in the main part, even an exponential amount of qubits can be saved for systems with low particle number, but at the expense of complicated gates.   For this Appendix, we firstly recap the situation of Section \ref{subsubsec:binadressing} and clarify what  binary addressing means.  Firstly, some nomenclature is introduced.
We then generalize the concept of binary addressing codes to weight-$K$ codes, using results from \cite{tian2007constant}. As an example, we explicitly obtain the $K=2$ code. \\
 Suppose we have a system with $N=2^r$ orbitals, and one particle in it.  Our goal is to encode the basis state, where the particle is on orbital $y\in [2^r]$, as a binary number in $r$ qubits. In this way, the state with occupational vector $\bbs{u_y}$ is encoded as $\ket{\bbs{q^{y,r}}}$, with $\bbs{q^{y,r}}\in\mathbb{Z}_2^{\otimes r}$ and $y=\mathrm{bin}( \bbs{q^{y,r}})+1$. Probing an unknown basis state, a decoding will now have components of the form
 \begin{align}\label{eq:binaddress1}
 \bbs{\omega} \to \prod_{i\in [r]} (\omega_i+ q^{y,r}_i+1 )\moto \, .
 \end{align} Such binary functions output $1$ only when $\bbs{\omega}=\bbs{q^{y,r}}$. In our nomenclature, we say that in the basis state $\ket{\bbs{q^{y,r}}}$, the particle has the coordinate $y$. We refer to codes that store particle coordinates in binary form,  as binary addressing codes.
\\

In the $K=1$ case from the main part, the code words just contain  the binary representation of one coordinate. The question is now how to generalize the binary addressing codes. For multi-weight codes, we have to have $K$ sub-registers  to store the addresses of $K$ particles. Naively, one would  want to store the coordinate of each particle in its respective sub-register in binary form, as we have done for $K=1$.  This however,  holds a problem. As the particles are indistinguishable, the stored coordinates would be interchangeable, the code would not be one-to-one. For the binary numbers $\bbs{\omega}^{\,1}$ and $\bbs{\omega}^{\,2}$, that represent a coordinate each, this would mean $\bbs{d}(\bbs{\omega}^{\,1}\oplus \bbs{\omega}^{\,2})=\bbs{d}(\bbs{\omega}^{\,2}\oplus \bbs{\omega}^{\,1})$. That strategy not only complicates the operator transform, it also leads to a certain qubit overhead, as each plain word has as many code words as there are permutations of $K$ items.  Since this naive idea leaves us unconvinced, we abandon it and search for one-to-one codes instead.  The key is to consider the coordinates to be in a certain format and this is where \cite{tian2007constant} comes into play. We proceed by using some relevant concepts of that paper. \\
 Let us consider the coordinates of $K$ particles to be given in the $N$-ary vector $\bbs{x}=(x_1, \dots , x_K)$. Between those coordinates, we have imposed an ordering $x_i>x_j$ as $i>j$.  Particles cannot share the same orbital, so we are excluding the cases where two coordinates are equal. Using results from \cite{tian2007constant}, we transform the latter into coordinates that  lack such an ordering, and where each component is an integer from a different range:
\begin{align}
\bbs{x}\;\to \;\bbs{y}=(y_1, \dots, y_K)^\top \qquad\text{with}\quad \bbs{y}\in \bigotimes_{m=1}^{K} \left[ \left\lceil \frac{N}{m}\right\rceil \right]\, .
\end{align}
Through that transform, each vector $\bbs{y}$ corresponds to a  valid  vector $\bbs{x}$, and there is no duplication.
We now represent the $\bbs{y}$-coordinates by binary numbers in the code words $\bbs{\omega}\in \mathbb{Z}_2^{\otimes n}$, where $n=\sum_{m=1}^K\left \lceil\log \frac{N}{m} \right\rceil $:
\begin{align} \label{eq:brick}
\bbs{\omega}=\bigoplus_{m=1}^{K}\bbs{q^{\, y_m,\left\lceil \frac{N}{m} \right\rceil}} \qquad \text{with} \quad \bbs{q^{i,j}} \in \mathbb{Z}_2^{\otimes j}\quad \text{and}\quad \mathrm{bin}\left( \bbs{q^{i,j}} \right)+1=i \, .
\end{align}
A geometric interpretation of the process portrays the vector $\bbs{x}$ as a set of coordinates in a $K$-dimensional, discrete vector space. The vectors allowed by the ordering form thereby a multi-dimensional tetrahedron. The states outside the tetrahedron do not correspond to a valid $\mathcal{V}$ vector, so encoding each coordinate $x_i$ in $\lceil \log N \rceil$ qubits would be redundant. We therefore dissect the tetrahedron,  and rearrange it into a \textit{brick}, as it is referred to in \cite{tian2007constant}. What is actually done is to apply symmetry operations (like point-reflections) on the vector space until the tetrahedron is deformed into the desired shape a $K$-dimensional, rectangular volume. The fact that the vectors to encode are now all inside a hyper-rectangle is what we wanted to achieve. We  can now clip the ranges of the coordinate axes (to $[\lceil \log\frac{N}{m} \rceil]$) to exclude vectors the vectors outside the brick. As the values on the axes correspond to non-binary addresses, this means that the qubit space is trimmed as well, and we have eliminated all states based on not-allowed coordinates. This is where we now reconnect to our task of finding a code: the $\bbs{e}$- and $\bbs{d}$-functions have to take into account the reshaping process, as only the coordinates $\bbs{x}$ have a physical interpretation and can be decoded. The binary addresses in the code words, on the other hand, are representatives of $\bbs{y}$. With binary logic, the two coordinates have to be reconnected.  We illustrate this abstract process on the example of the $(K=2)$-code.
\subsection*{Weight-two binary addressing code }
As an example, we present the weight-two binary addressing code on $N=2^r$ orbitals. The integer $r$ will determine the size of the entire qubit system $n=2r-1$, with two registers of size $r$ and $r-1$.  \\
\begin{figure}
\begin{tikzpicture}[scale=0.5]
\path[fill=black] (0,8)--(7,8)--(7,7)--(6,7)--(6,6)--(5,6)--(5,5)--(4,5)--(4,4)--(0,4)--cycle;
\path[fill=black!22] (8,7)--(8,4)--(5,4)--(5,5)--(6,5)--(6,6)--(7,6)--(7,7)--cycle;
\path[fill = black!22] (0,1)--(1,1)--(1,2)--(2,2)--(2,3)--(3,3)--(3,4)--(4,4)--(0,4)-- cycle ;
\draw[](1,0)--(1,8);
\draw[](2,0)--(2,8);
\draw[](3,0)--(3,8);
\draw[](4,0)--(4,8);
\draw[](5,0)--(5,8);
\draw[](6,0)--(6,8);
\draw[](7,0)--(7,8);
\draw[](8,0)--(8,8);
\draw[](0,0)--(8,0);
\draw[](0,1)--(8,1);
\draw[](0,2)--(8,2);
\draw[](0,3)--(8,3);
\draw[](0,5)--(8,5);
\draw[](0,6)--(8,6);
\draw[](0,7)--(8,7);
\draw[ color=white] (0,4)--(0,8);
\draw[ color=white] (1,4)--(1,8);
\draw[ color=white] (2,4)--(2,8);
\draw[ color=white] (3,4)--(3,8);
\draw[ color=white] (4,4)--(4,8);
\draw[ color=white] (5,5)--(5,8);
\draw[ color=white] (6,6)--(6,8);
\draw[ color=white] (7,7)--(7,8);
\draw[ color=white] (0,8)--(7,8);
\draw[ color=white] (0,7)--(7,7);
\draw[ color=white] (0,6)--(6,6);
\draw[ color=white] (0,5)--(5,5);
\draw[ color=white] (0,4)--(4,4);
\draw[](0,4)--(8,4);
\draw[](0,0)--(0,8);
\draw[](0,8)--(8,8);
\node[left] at (0,0.5) {1};
\node[left] at (0,1.5) {2};
\node[left] at (0,3.5) {$N/2$};
\node[left] at (0,4.5) {$1+N/2$};
\node[left] at (0,7.5) {$N$};
\node[left] at (-3.5,4) {$x_2$};
\draw[|->|] (-3.5,0)--(-3.5,8);
\node[right] at (8,4.5) {1};
\node[right] at (8,5.5) {2};
\node[right] at (8,7.5) {$N/2$};
\node[right] at (10,6) {$y_2$};
\draw[|->|] (10,4)--(10,8);
\node[below] at (.5,0) {1};
\node[below] at (1.5,0) {2};
\node[below] at (3.5,0) {$\frac{N}{2}$};
\node[below] at (4.5,0) {$ $};
\node[below] at (6.5,0) {$ $};
\node[below] at (7.5,0) {$N$};
\draw[|->|] (0,-2)--(8,-2);
\node[below] at (4,-2) {$x_1 = y_1$ };
\draw[ultra thick] (0,0)--(8,8);
\end{tikzpicture}
\caption{Visualization of the 2-dimensional vector space: a valid vector is represented as a colored tile. The left gray tiles and the black ones constitute the triangle, defining all valid vectors $\bbs{x}=(x_1, x_2)^\top$. The marked diagonal tiles are to be excluded from the encoded space. The black tiles and the gray ones on the right of this diagonal form the brick,  containing all $\bbs{y}=(y_1, y_2)^\top$ vectors.  }
\label{fig:triangle}
\end{figure}
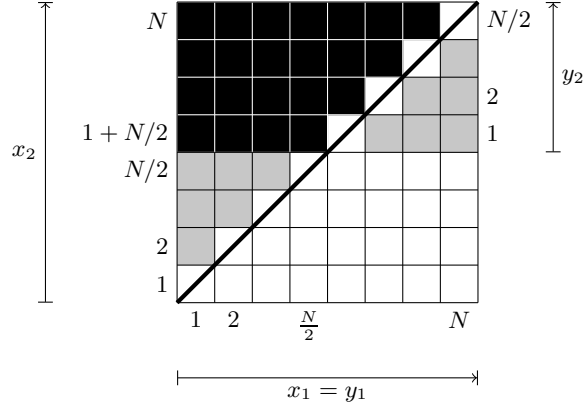
With the two registers, a binary vector $\bbs{\omega}=\bbs{\alpha}\oplus\bbs{\beta}$ with $\bbs{\alpha}\in \mathbb{Z}_2^{\otimes r}$ and $\bbs{\beta}\in \mathbb{Z}_2^{\otimes (r-1)}$ is defining the qubit basis. In two dimensions, the brick turns into a rectangle and the tetrahedron into triangle. The decoding function takes binary addresses of the rectangular $\bbs{y}$, and transforms them into coordinates in the triangle $\bbs{x}$. The ordering condition implies hereby where to dissect the rectangle: Figure \ref{fig:triangle} may serve as a visual aid, disregarding the excluded cases of $y_1=y_2$, we find for $y_1\in [N]$, $y_2\in [N/2]$ and $\bbs{x} \in [N]^{\otimes 2}$:
\begin{align}
(x_1,x_2) = \begin{cases} (y_1,N/2+y_2)& \text{for}\quad  y_1<N/2+y_2  \\ (y_1,N/2-y_2+1)& \text{for}\quad y_1>N/2+y_2 \, . \end{cases}
\end{align}
This decoding is translated into a binary functions as follows: the coordinate $y_1$ is represented by the binary vector $\bbs{\alpha}$ and $y_2 $ by $\bbs{\beta}$. For each component defined by the binary vector $\bbs{b}\in \mathbb{Z}_2^{\otimes r}$, we have
\begin{align}
d_{j}\left( \bbs{\alpha}\oplus \bbs{\beta}\right)\; =\; &S\left( \bbs{\alpha}, \bbs{\beta} \right) \prod_{i=1}^{r} \left(\alpha_i+q^{\,j,r}_i+1 \right)+\left(1+S\left( \bbs{\alpha}, \bbs{\beta} \right)\right)\left(1+T\left( \bbs{\alpha}, \bbs{\beta} \right)\right)\prod_{i=1}^{r} \left(\alpha_i+q^{\,j,r}_i \right)\nonumber \\&+\left(1+S\left( \bbs{\alpha}, \bbs{\beta} \right)\right)\left(1+T\left( \bbs{\alpha}, \bbs{\beta} \right)\right)\prod_{k=1}^{r-1} \left( \beta_k+q^{\,j,r}_k \right)+ S\left( \bbs{\alpha}, \bbs{\beta} \right) \prod_{k=1}^{r-1} \left( \beta_k+q^{\,j,r}_k+1 \right) \;\moto\, ,
\end{align}
with $\bbs{q^{\,j,r}}=(q^{\,j,r}_1, q^{\,j,r}_2, \dots, q^{\,j,r}_{2^r})$ as defined in \eqref{eq:brick} and we have employed two binary functions $S$ and $T:(\mathbb{Z}_2^{\otimes r}, \mathbb{Z}_2^{\otimes (r-1)})\to \mathbb{Z}_2 $. Here, $S$ compares the binary numbers to determine if the coordinates are left of the dissection (a black tile in Figure \ref{fig:triangle}).
\begin{align}
S\left( \bbs{\alpha}, \bbs{\beta} \right)=\alpha_r\sum_{j=1}^{r-1} \left[ \prod_{r-1\geq i > j} \left( \alpha_i+\beta_i+1 \right) \right](1+\alpha_j)\beta_j+1+\alpha_r \; \moto
\end{align}
The binary function $T$, on the other hand, is  checking whether a set of coordinates is on a diagonal position (diagonally marked tiles). These excluded cases are mapped to $\left( 0\right)^{\otimes r}$ altogether.
\begin{align}
T\left( \bbs{\alpha},\bbs{\beta}\right)=\prod_{i}\left(\alpha_i+\beta_i \right)
\moto \end{align}
This concludes the decoding function. Unfortunately, the amount of logic elements in the decoding will complicate the weight-two codes quite a bit, and the encoding function is hardly better. The reason for this is to find in the ordering condition: the update operations are conditional on whether we change the ordering of the coordinates represented by $\bbs{\alpha}$ and $\bbs{\beta}$. This is reflected in a non-linear encoding function: we remind us that the encoding function is a map $\bbs{e}:\mathbb{Z}_2^{\otimes 2^r} \to \mathbb{Z}_2^{\otimes (2r-1)}$, and with $\bbs{\nu} \in \mathbb{Z}_2^{\otimes 2^r}$ we find
\begin{align}
\bbs{e}\left( \bbs{\nu}\right) \quad=\qquad \sum_{j=2}^{2^{r-1}}\; \sum_{i=1}^{j-1}\left(\bbs{q^{\, i,r}}+\bbs{I^{\,r}} \moto \right) &\oplus \left(\bbs{q^{\, j,r-1}}+\bbs{I^{\,r-1}}\moto \right) \nu_i \nu_j \nonumber \\ +\sum_{j=2^{r-1}+1}^{2^{r}} \; \sum_{i=1}^{2^{r-1}}\left(\bbs{q^{\, i,r}}+\bbs{I^{\,r}} \moto \right) &\oplus \left(\bbs{q^{\, j-2^{r-1},r-1}} \right) \nu_i \nu_j \nonumber  \\ +\sum_{j=2^{r-1}+2}^{2^{r}} \;\sum_{i=2^{r-1}+1}^{j-1}\left(\bbs{q^{\, i,r}}\right) &\oplus \left(\bbs{q^{\, j-2^{r-1},r-1}}\right) \nu_i \nu_j \nonumber \quad \moto \, ,
\end{align}
with $\bbs{q^{\, i,j}}$ as defined in \eqref{eq:brick}, and $\bbs{I^{\,j}}=\left( 1 \right)^{\otimes j}=\bbs{q^{ 2^j\!, \,j}}$. \\
The dissecting of tetrahedrons can be generalized for codes of weight larger than two (see again\cite{tian2007constant}), but as one increases the number of dissections, the code functions are complicated even further.

\section{Segment codes}
\label{sec:ments}
In this Appendix, we provide detailed information on the segment codes. We firstly concern ourselves with the segmentation of the global code, including a derivation of the segment sizes. In another subsection we  construct the segment codes themselves. The last subsection is dedicated to the adjustments one has to make to  Hamiltonian, such that segment codes become feasible to use.
\subsection{Segment sizes}
At this point we want to sketch the idea behind the segment sizes ($\hat{N}$, $\hat{n}$) stated during  Section \ref{subsubsec:mentcodes} in the main part, but first of all we would like to clearly set up the situation.

We consider vectors  $\bbs{\nu}\in\mathbb{Z}_2^{\otimes N}$ to consist of $\hat{m}$ smaller vectors $\bbs{\hat{\nu}^i}$ of length $\hat{n}+1$, such that $\bbs{\nu}=\bigoplus_{i=1}^{\hat{m}} \bbs{\hat{\nu}^i}$. We call those vectors $\bbs{\hat{\nu}^i}$ segments of $\bbs{\nu}$. The goal is now to find a code ($\bbs{e}$, $\bbs{d}$) to encode a basis $\mathcal{V}$ which contains all vectors $\bbs{\nu}$ with Hamming weight $K$. For that purpose we relate the segment $\bbs{\hat{\nu}^i}$  to a segment  of the code space, $\bbs{\hat{\omega}^i}$, for all $i\in [N]$. The code space segments constitute the code words in a fashion similar to the previous segmentation of $\bbs{\nu}$: $\bbs{\omega}=\bigoplus_{i=1}^{\hat{m}} \bbs{\hat{\omega}^i}$. However, the length of those binary vectors   $\bbs{\hat{\omega}^i}$ is  $\hat{n}$, such that with $n=\hat{m}\hat{n}$ and $N=\hat{m}(\hat{n}+1)$, the problem is reduced by $\hat{m}$ qubits as compared to conventional transforms.  We now introduce the \textit{subcodes}  ($\bbs{\hat{e}}:\mathbb{Z}_2^{\otimes (\hat{n}+1)} \to \mathbb{Z}_2^{\otimes \hat{n}},\bbs{\hat{d}}:\mathbb{Z}_2^{\otimes \hat{n}} \to \mathbb{Z}_2^{\otimes( \hat{n}+1)}$), with which we encode the $i$-th segment $\bbs{\hat{\nu}^i}$ as $\bbs{\hat{\omega}^i}$ (see Figure \ref{fig:sgmnts}). Note that we require the subcodes to inherit all the code properties. In this way we guarantee the code properties of the \textit{global code} ($\bbs{e}$, $\bbs{d}$) when appending $\hat{m}$ instances of the same subcode:

\begin{align}
\label{eq:segmentcodes}
\bbs{d}\left( \bigoplus_{i=1}^{\hat{m}} \bbs{\hat{\omega}^i} \right)=\bigoplus_{i=1}^{\hat{m}} \bbs{\hat{d}} \left( \bbs{\hat{\omega}^i}  \right) \quad , \qquad  \bbs{e}\left( \bigoplus_{i=1}^{\hat{m}} \bbs{\hat{\nu}^i} \right)=\bigoplus_{i=1}^{\hat{m}} \bbs{\hat{e}} \left( \bbs{\hat{\nu}^i}  \right)\, .
\end{align}

\newcommand{\dist}{2.7}
\newcommand{\arrowdist}{0.1}
\begin{figure}
\begin{tikzpicture}[scale=.7]
\node[above] at (2.9, .6) {$\bbs{\hat{\nu}^1}$};
\node[above] at (7.9, .6) {$\bbs{\hat{\nu}^2}$};
\node[above] at (17.9, .6) {$\bbs{\hat{\nu}^{\hat{m}}}$};
\draw[<-] (-.45-\arrowdist,-.75*\dist)--(-.45-\arrowdist,-.25*\dist);
\draw[->] (-.45+\arrowdist,-.75*\dist)--(-.45+\arrowdist,-.25*\dist);
\node [left] at (-.45-\arrowdist,-.5*\dist) {$\bbs{e}$};
\node [right] at (-.45+\arrowdist,-.5*\dist) {$\bbs{d}$};
\draw[<-] (2.9-\arrowdist,-.75*\dist)--(2.9-\arrowdist,-.25*\dist);
\draw[->] (2.9+\arrowdist,-.75*\dist)--(2.9+\arrowdist,-.25*\dist);
\node [left] at (2.9-\arrowdist,-.5*\dist) {$\bbs{\hat{e}}$};
\node [right] at (2.9+\arrowdist,-.5*\dist) {$\bbs{\hat{d}}$};
\node [left] at (7.9-\arrowdist,-.5*\dist) {$\bbs{\hat{e}}$};
\node [right] at (7.9+\arrowdist,-.5*\dist) {$\bbs{\hat{d}}$};
\node [left] at (17.9-\arrowdist,-.5*\dist) {$\bbs{\hat{e}}$};
\node [right] at (17.9+\arrowdist,-.5*\dist) {$\bbs{\hat{d}}$};
\draw[<-] (7.9-\arrowdist,-.75*\dist)--(7.9-\arrowdist,-.25*\dist);
\draw[->] (7.9+\arrowdist,-.75*\dist)--(7.9+\arrowdist,-.25*\dist);
\draw[<-] (17.9-\arrowdist,-.75*\dist)--(17.9-\arrowdist,-.25*\dist);
\draw[->] (17.9+\arrowdist,-.75*\dist)--(17.9+\arrowdist,-.25*\dist);

\node at (-.5,.1) {$\bbs{\nu} =$};
\node at (0,0) {  \Huge $\;\;($};

\node at (1,0) {$\;\;\hat{\nu}^1_1,$};
\node at (2,0) {$\hat{\nu}^1_2,$};
\node at (3,0) {$\hat{\nu}^1_3,$};
\node at(4,0) {$\hat{\nu}^1_4,$};
\node at (5,0) {$\hat{\nu}^1_5$};
\node[below] at (5.5,0) {$,$};
\node at (6,0) {$\;\;\hat{\nu}^2_1,$};
\node at(7,0) {$\hat{\nu}^2_2,$};
\node at (8,0) {$\hat{\nu}^2_3,$};
\node at (9,0) {$\hat{\nu}^2_4,$};
\node at (10,0) {$\hat{\nu}^2_5$};
\node[below] at (10.5,0) {$,$};
\node at (16,0) {$\;\;\hat{\nu}^{\hat{m}}_1,$};
\node at (17,0) {$\hat{\nu}^{\hat{m}}_2,$};
\node at (18,0) {$\hat{\nu}^{\hat{m}}_3,$};
\node at (19,0) {$\hat{\nu}^{\hat{m}}_4,$};
\node at (20,0) {$\hat{\nu}^{\hat{m}}_5$};
\node at(20.5,0){ \Huge $\,)$};
\node[] at (13,0) {\Huge $\cdots$};
\draw[ultra thick] (.6,.5)--(5.4,.5)--(5.4,-.5)--(.6,-.5)--cycle;
\draw[ultra thick] (5.6,.5)--(10.4,.5)--(10.4,-.5)--(5.6,-.5)--cycle;
\draw[ultra thick] (15.6,.5)--(20.4,.5)--(20.4,-.5)--(15.6,-.5)--cycle;
\node at (-.5,-\dist+.1) {$\bbs{\omega} = $};
\node at (0,-\dist)  {\Huge $\;\;($};
\node at (2.9,-\dist) {$\hat{\omega}^1_1$, $\;\;\hat{\omega}^1_2$, $\;\;\hat{\omega}^1_3$, $\;\;\hat{\omega}^1_4$};
\node[below] at (2.9, -\dist-.6) {$\bbs{\hat{\omega}^1}$};
\node[below] at (7.9, -\dist-.6) {$\bbs{\hat{\omega}^2}$};
\node[below] at (17.9, -\dist-.6) {$\bbs{\hat{\omega}^{\hat{m}}}$};
\node at (7.9,-\dist) {$\hat{\omega}^2_1$, $\;\;\hat{\omega}^2_2$, $\;\;\hat{\omega}^2_3$, $\;\;\hat{\omega}^2_4$};
\node at (18.1,-\dist) {$\hat{\omega}^{\hat{m}}_1$, $\;\;\hat{\omega}^{\hat{m}}_2$, $\;\;\hat{\omega}^{\hat{m}}_3$, $\;\;\hat{\omega}^{\hat{m}}_4$};
\node[below] at (5.5,-\dist) {$,$};
\node[below] at (10.5,-\dist) {$,$};
\node at(20.5,-\dist){\Huge $\,)$};
\node[] at (13,-\dist) {\Huge $\cdots$};
\draw[ultra thick] (.6,.5-\dist)--(5.4,.5-\dist)--(5.4,-.5-\dist)--(.6,-.5-\dist)--cycle;
\draw[ultra thick] (5.6,.5-\dist)--(10.4,.5-\dist)--(10.4,-.5-\dist)--(5.6,-.5-\dist)--cycle;
\draw[ultra thick] (15.6,.5-\dist)--(20.4,.5-\dist)--(20.4,-.5-\dist)--(15.6,-.5-\dist)--cycle;
\end{tikzpicture}
\caption{Visualization of \eqref{eq:segmentcodes} for $\hat{n}=4$. The global code ($\bbs{e}$, $\bbs{d}$) relates the occupation vectors to the global code words $\bbs{\nu} \leftrightarrow \bbs{\omega}$. The an instance of the subcode ($\bbs{\hat{e}}$, $\bbs{\hat{d}}$) relates  $i$-th block in $\bbs{\nu}$, $\bbs{\hat{\nu}^i}$, to the $i$-th segment in the code words, $\bbs{\hat{\omega}^i}$.}
\label{fig:sgmnts}
\end{figure}
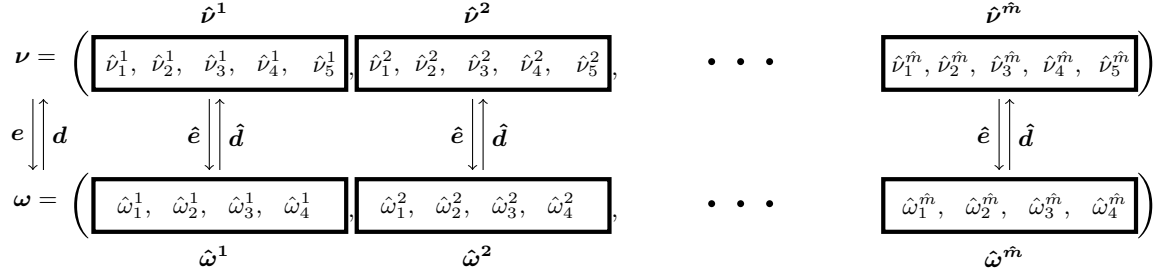
The  orbital number being an integer multiple of the block size is of course an idealized scenario. One will probably have to add a few other components  in order to compensate for dimensional mismatches.  \\
We now set out to find the smallest segment size $\hat{n}$. It should be clear that  $\hat{n}$ is a function of the targeted Hamming weight $K$:  this means $K$ determines which segment codes  are suitable for the system. The reason for this is that we need to encode all vectors with weight $0$ to  $K$ inside every segment, taking into account for the up to $K$ particles on the orbitals inside one segment. In order to include weight-$K$ vectors, the size of each segment must be at least $K$. If the segment size would be exactly $K$, on the other hand, we end up encoding the entire Fock space again. In doing so, we are not making any qubit savings.  The segments must thus be larger than $K$.  In other words, we look for an integer $\hat{n}>K$, where the sum of all combinations $\bbs{\hat{\nu}} \in \mathbb{Z}_2^{\otimes (\hat{n}+1)}$ with $\hamming{\bbs{\hat{\nu}}}\leq K$ is smaller equal $2^{\hat{n}}$.
\begin{align}
2^{\hat{n}}\; \geq \; \sum_{k=0}^{K} \binom{\hat{n}+1}{k}
\end{align}
In the case $\hat{n}=2K$, the condition is fulfilled as identity, since exactly half of all $2^{\hat{n}+1}$ combinations are included in the sum.
\subsection{Subcodes}
This subsection offers a closer look at the construction of the segment subcodes ($\bbs{\hat{e}}$, $\bbs{\hat{d}}$). Let us start by considering  the decoding $\bbs{\hat{d}}$ in order to explore the nature of the binary switch $f(\bbs{\hat{\omega}})$, that occurs in \eqref{eq:segdec}.
One observes the two (affine) linear  $\left(\mathbb{Z}_2^{\otimes \hat{n}} \to \mathbb{Z}_2^{\otimes ( \hat{n}+1)}\right)$-maps
\begin{align}
\label{eq:halves}
\bbs{\hat{\omega}} \;  \to \; \left[\begin{matrix}
1 \\
& \ddots \\
& & 1 \\
0 & \dots & 0
\end{matrix} \right]\boldsymbol{\hat{\omega}} \mod 2 \quad , \qquad  \bbs{\hat{\omega}} \; \to \;\left[\begin{matrix}
1 \\
& \ddots \\
& & 1 \\
0 & \dots & 0
\end{matrix} \right]\boldsymbol{\hat{\omega}} \;+\;\left( \begin{matrix}
1 \\ \vdots \\ \vdots \\1
\end{matrix} \right) \moto \, .
\end{align}
to produce together all the vectors with weight equal or smaller than $K$, if we input all $\bbs{\hat{\omega}}$ with $\hamming{\bbs{\hat{\omega}}}\leq K$ into the first, and the remaining cases with $\hamming{\bbs{\hat{\omega}}}> K$ into the second one. Note that the last component is always zero in outputs of the first function and one in the second. Therefore, the inverse of both maps is always a linear map with the  matrix $\left[ \;\mathbb{I}\; \right| \left. \bbs{I^{\,\hat{n}}} \right]$. We take this inverse as encoding \eqref{eq:segenc}, and the two maps \eqref{eq:halves} are merged into the decoding \eqref{eq:segdec}.
In order to switch between these two maps we define the binary function $f(\bbs{\hat{\omega}}): \mathbb{Z}_2^{\otimes\hat{n}}\to \mathbb{Z}_2$ such that
\begin{align}
f(\bbs{\hat{\omega}})=\begin{cases} 1 & \text{for}\;\; \hamming{\bbs{\hat{\omega}}}>K \\ 0 & \text{otherwise}\; . \end{cases}
\end{align}
In general, one can define this binary switch in a brute-force way by
 \begin{align}
f\left( \bbs{\hat{\omega}} \right) = \sum_{k=K+1}^{2K} \sum_{\begin{smallmatrix} \bbs{t}\in \mathbb{Z}_2^{\otimes 2K} \\ \hamming{\bbs{t}}=k \end{smallmatrix}} \prod_{m=1}^{2K} \left(\hat{\omega}_m+1+t_m \right)\moto \, .
\end{align}

For the case $K=1$ ($\hat{n}=2$), the switch equals $f(\bbs{\omega})=\omega_1 \omega_2$, and for the code we recover a version of  binary addressing codes, where the vector $(0,0,0)$ is encoded.
\begin{align}
\bbs{\hat{d}}\left(\bbs{\hat{\omega}}\right)= \left( \begin{matrix}
\hat{\omega}_1\,(\hat{\omega}_2+1) \\
(\hat{\omega}_1+1)\,\hat{\omega}_2 \\
\hat{\omega}_1 \hat{\omega}_2
\end{matrix}\right) \moto\quad  , \qquad  \bbs{\hat{e}}\left(\bbs{\hat{\nu}} \right) = \left[ \begin{matrix}
1 & 0& 1 \\
0 & 1 & 1
\end{matrix}\right] \bbs{\hat{\nu}}\;\moto\; .
\end{align}
In the $K=2$ ($\hat{n}=4$) case, this binary switch is found to be $f\left( \bbs{\hat{\omega}} \right)=\hat{\omega}_1\hat{\omega}_2\hat{\omega}_3+\hat{\omega}_1\hat{\omega}_2\hat{\omega}_4+\hat{\omega}_1\hat{\omega}_3\hat{\omega}_4+\hat{\omega}_2\hat{\omega}_3\hat{\omega}_4+\hat{\omega}_1\hat{\omega}_2\hat{\omega}_3\hat{\omega}_4 \moto$.

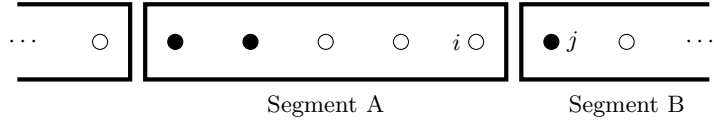
\begin{figure}
\begin{tikzpicture}[scale=1]
\draw[] (0,0) circle[radius=0.1];
\draw[fill=black] (1,0) circle[radius=0.1];
\draw[fill=black] (2,0) circle[radius=0.1];
\draw[] (3,0) circle[radius=0.1];
\draw[] (4,0) circle[radius=0.1];
\draw[] (5,0) circle[radius=0.1];
\draw[fill=black] (6,0) circle[radius=0.1];
\draw[]	(7,0) circle[radius=0.1];
\node[] at (8,0) {$\cdots$};
\node[] at (-1,0) {$\cdots$};
\node[left] at (5,0) {$i\;$};
\node[ right] at (6,0) {$\; j$};
\draw[ultra thick] (.6,.5)--(5.4,.5)--(5.4,-.5)--(.6,-.5)--cycle;
\draw[ultra thick] (-1.1,.5)--(.4,.5)--(.4, -.5)--(-1.1,.-.5);
\draw[ultra thick] (8.1,.5)--(5.6,.5)--(5.6,-.5)--(8.1,-.5);
\node[below] at (3,-.6) {Segment A};
\node[below] at (7,-.6) {Segment B};
\end{tikzpicture}
\caption{(Filled) Circles represent (occupied) fermionic orbitals, where $K=2$ segment codes are used in the indicated blocks.  This occupational case is problematic for the codes, as the operator $c^{\dagger}_i c^{\;}_j$ acting on this state leaves the encoded space. }
\label{fig:blocks}
\end{figure}

\subsection{Hamiltonian adjustments}
As mentioned in Section \ref{subsubsec:mentcodes}, in the main part, segment codes are not automatically compatible with all particle-number-conserving Hamiltonians. We show here, how certain adjustments can be made to these Hamiltonians, such that their action on the space $\mathcal{H}_N^K$ is not changed, but segment codes become feasible to describe them with. In order to understand this issue, we begin by examining the encoded space. For that purpose we reprise the situation of \eqref{eq:segmentcodes}, where we have append $\hat{m}$ instances of the same subcode.
With segment codes, the basis $\mathcal{V}$ contains vectors with Hamming weights from $0$ to $\hat{m}K$. We have encoded all possible vectors $\bbs{\nu}$ with $0\leq\hamming{\bbs{\nu}}\leq K$, but although we have some, not all vectors with $\hamming{\bbs{\nu}}>K$ are encoded. We can illustrate that point rather quickly: each segment has length $2K+1$, but the subcode encodes vectors $\bbs{\hat{\nu}}$ with only $\hamming{\bbs{\hat{\nu}}}\leq K$. The (global) basis $\mathcal{V}$ is thus deprived of vectors $\bbs{\nu}=(\bigoplus_i \bbs{\hat{\nu}^i})$ where for any segment $i$, $\hamming{\bbs{\hat{\nu}^i}}>K$. \\
We now turn our attention to terms, which, when present in a Hamiltonian, make segment codes infeasible to use.
Note, that $\mathcal{V}$-vectors with $\hamming{\bbs{\nu}}\neq K$, are not corresponding to fermionic states we are interested in.  In particular it is a certain subset of states with $\hamming{\bbs{\nu}}> K$, which can lead out of the encoded space (into the states previously mentioned) when acted upon with certain fermionic operators. Let us consider  the operator $c^{\dagger}_i c^{\;}_j$ as an example, where $i$ and $j$ are in different segments (let us call these segments A and B). Now a  basis state  as depicted in Figure \ref{fig:blocks}, is not annihilated by $c^{\dagger}_i c^{\;}_j$, and leads into a state with $3$ particles in segment A. The problem is that the initial state is encoded in the $(K=2)$ segment codes, whereas the updated state (with the 3 particles in A) is not. In general, operators $\hat{h}_{\bbs{ab}}$, that change occupations in between segments, will cause some basis states with $\hamming{\bbs{\nu}}>K$ to leave the encoded space. We can however adjust these terms $\hat{h}_{\bbs{ab}}\to\hat{h}^\prime_{\bbs{ab}}$,  such that  $\hat{h}^\prime_{\bbs{ab}}:\mathrm{span}(\mathcal{B}) \to \mathrm{span}(\mathcal{B})$, where $\mathcal{B}$ is the basis encoded by the segment codes. We now sketch the idea behind those adjustments, before we reconsider the situation of Figure \ref{fig:blocks}. Note that after these adjustments have been made to all Hamiltonian terms in question, the segment codes are compatible with the new Hamiltonian.
The  idea is to switch those terms off for states, that already have $K$ particles inside the segments, to which particles will be added. We have to take care to do this in a way that leaves the Hamiltonian hermitian on the level of second quantization, i.e. we have to adjust the terms $\hat{h}^{\;}_{\bbs{ab}}$ and  $\hat{h}^\dagger_{\bbs{ab}}$ into $\hat{h}^{\prime}_{\bbs{ab}}$ and $(\hat{h}^\dagger_{\bbs{ab}})^\prime$, such that $\hat{h}^{\prime}_{\bbs{ab}}+(\hat{h}^\dagger_{\bbs{ab}})^\prime$ is hermitian. 
For the $K=2$ code of Figure \ref{fig:blocks}, we can make the following adjustments:

\begin{align}
c^{\dagger}_i c^{\;}_j \; \to \; \left(1- \sum_{\mathclap{ l,k<l\, \in \, \text{ B}}} c^{\dagger}_k c^{\;}_k c^{\dagger}_l c^{\;}_l \right)c^{\dagger}_i c^{\;}_j \left(1-\sum_{\mathclap{ w,v<w \,\in \, \text{ A}}} c^{\dagger}_v c^{\;}_v c^{\dagger}_w c^{\;}_w \right)\, .
\end{align}

\section{Conventional mappings}
\label{sec:linear}
We now revisit the  conventional transforms from Section \ref{sec:background} in the main part,  and discuss all notations that have been introduced to express it close to the appealing nomenclature of  \cite{seeley2012bravyi,tranter2015bravyi}.
In particular, we show that the relation \eqref{eq:line} is recovered as a special case from \eqref{eq:admiral} and \eqref{eq:linenc}. After that,  we verify that such  constructions satisfy the fermionic anticommutation relations.  For now, however, we would like to restate the situation: a linear  $n=N$ code, encoding the entire Fock space,  is mediated by the quadratic matrices $A$ and $A^{-1}$, such that $\bbs{e}\left(\bbs{\nu}\right)=(A\bbs{\nu}\moto)$ and $\bbs{d}\left( \bbs{\omega} \right)=(A^{-1}\bbs{\omega}\moto)$. The matrices are required to be each others inverses, so
\begin{align}
\label{eq:inverse}
\sum_{j=1}^N A_{ij}\;(A^{-1})_{jk}\moto=\delta_{ik}\, .
\end{align}
We now explain the form of the parity, update and flip sets. As the code is linear, the extraction operator is retrieving only Pauli strings following \eqref{eq:prop2} and \eqref{eq:prop4}.
One finds:
\begin{align}
\label{eq:sets0}
&\mathfrak{X}\left[ d_i \right]= \mathfrak{X}\left[ \bbs{\omega}\to \sum_j\left( A^{-1} \right)_{ij}\omega_j\moto \right] = \bigotimes_{j\in[N]} \left(Z_j\right)^{\left( A^{-1} \right)_{ij}} \;=\; \bigotimes_{j\in F(i)} Z_j \\ \label{eq:sets1}
&\mathfrak{X}\left[ p_i \right]= \mathfrak{X}\left[ \bbs{\omega}\to \sum_{j<i}\sum_k\left( A^{-1} \right)_{jk}\omega_k\moto \right] = \bigotimes_{k\in[N]} \left(Z_j\right)^{\left( R A^{-1} \right)_{ik}} \;=\; \bigotimes_{k\in P(i)} Z_k \, ,
\end{align}
where $P(i)$ and $F(i)$ are the parity and flip sets  with respect to $i$, as we defined them in Section \ref{sec:background}. The update sets $U(i)$ are obtained from update operators of linear encodings:
\begin{align}
\label{eq:sets3}
\mathcal{U^{\,\bbs{a}}}=\bigotimes_{i} \left(X_i\right)^{e_i\left(\bbs{q} \right)}=\bigotimes_{i} \left(X_i\right)^{\sum_j A_{ij}q_j} = \prod_{k\in[l]}\bigotimes_{i\in[N]} \left(X_i\right)^{ A_{ia_k}} = \prod_{k\in[l]}\;\bigotimes_{i\in U\left(a_k\right)} X_i \, .
\end{align}
In order to derive \eqref{eq:line}, we would like to point out the  commutation relations between Pauli strings $(\bigotimes_{u\in U(i)}X_u)$, $(\bigotimes_{v\in F(j)} Z_v)$ and $(\bigotimes_{w\in P(k)} Z_w)$.  These will prove useful in verifying the fermionic commutation relations later.
For commutations of update- and flip-set strings we find:
\begin{align}
 \label{eq:commute1}
&\left(\bigotimes_{u\in U(i)} X_u \right)\left( \bigotimes_{v\in F(j)} Z_v \right)= \bigotimes_{w\in[N]} \left(X_w \right)^{A_{iw}} \left(Z_w \right)^{(A^{-1})_{wj}} = \bigotimes_{w\in[N]} \left(-1 \right)^{A_{iw}( A^{-1})_{wj}} \left(Z_w \right)^{(A^{-1})_{wj}} \left(X_w \right)^{A_{iw}}\\
&=\left( -1\right)^{\sum_w A_{iu}( A^{-1})_{wj} }\left( \bigotimes_{v\in F(j)} Z_v \right)\left(\bigotimes_{u\in U(i)} X_u \right) = \left( -1\right)^{\delta_{ij}}\left( \bigotimes_{v\in F(j)} Z_v \right)\left(\bigotimes_{u\in U(i)} X_u \right) \, .
\end{align}
We have used the relation \eqref{eq:inverse} for the above. Similarly, for commutations of update and parity strings we have:
 \begin{align}
 \label{eq:commute2}
 &\left(\bigotimes_{u\in U(i)} X_u \right)\left( \bigotimes_{w\in P(j)} Z_w \right) = \left( -1\right)^{\theta_{ij}}\left( \bigotimes_{w\in P(j)} Z_w \right) \left(\bigotimes_{u\in U(i)} X_u \right)\, .
 \end{align}
Finally, we combine \eqref{eq:sets0}-\eqref{eq:sets3} with the operator from \eqref{eq:admiral}. Using \eqref{eq:commute1} and \eqref{eq:commute2} to move every update string $\left(\bigotimes_{u\in U(a_j)} X_u \right)$ in between the projectors and parity strings of $a_j$ and $a_{j+1}$, we get
 \begin{align}
 &\mathcal{U}^{\,\bbs{a}} \left( \prod_{v=1}^{l-1}\,\prod_{w=v+1}^{l}\left(-1 \right)^{\theta_{a_v a_w}} \right) \left(\prod_{x=1}^{l} \frac{1}{2}\left(\mathbb{I}- \left[\prod_{y=x+1}^{l}(-1)^{\delta_{a_x a_y}}\right](-1)^{b_x}\;\mathfrak{X}\left[d_{a_x}\right]\right) \mathfrak{X}\left[ \,p_{a_x} \right] \right) \\
 &= \prod_{x=1}^{l} \left( \frac{1}{2}\left(\bigotimes_{u\in U(a_x)} X_u\right)\left(\mathbb{I}-(-1)^{b_x} \bigotimes_{v\in F(a_x)} Z_v\right)  \bigotimes_{w\in P(a_x)} Z_w \right) \, ,
 \end{align}
 which is a sequence of the operators \eqref{eq:line}. The transform of a singular operator is $c_j^{(\dagger)}$ is thus derived from \eqref{eq:admiral}. Although we have  already shown that  \eqref{eq:admiral} satisfies \eqref{eq:stateaction1}-\eqref{eq:stateaction4}, but we now want to show that \eqref{eq:line} fulfills the anticommutation relations \eqref{eq:antiraw}  in particular.
In doing so, we generally distinguish the cases $i=j$ and $i\neq j$.
For $[c_j^{(\dagger)},c_j^{(\dagger)}]_{+}$, we consult \eqref{eq:commute1} and find
 \begin{align}
 c^\dagger_j c^\dagger_j \; = \; c^{\;}_j c^{\;}_j \;\hat{=}\; \frac{1}{4}\left( \mathbb{I}-\bigotimes_{v\in F(j)}Z_v\right)\left( \mathbb{I}+\bigotimes_{w\in F(j)}Z_w\right) \; = \; 0 \, .
 \end{align}
  We notice that for $i \neq j$,  the gate transform of $c^{\;}_i c^{\;}_j$ $\left( c^{\dagger}_i c^{\dagger}_j\right)$ properly differs by a minus sign from the transform of $c^{\;}_j c^{\;}_i$ $\left( c^{\dagger}_j c^{\dagger}_i\right)$ due to \eqref{eq:commute2}. We want to make this observation explicit for the  $i\neq j$ case of $[c_i^{},c_j^{\dagger}]_{+}$:

  \begin{align}
  &c^{\;}_i c^{\dagger}_j \; \hat{=} \; \frac{1}{4} \left(\bigotimes_{u\in U(i)} X_u\right)\left(\mathbb{I}- \bigotimes_{v\in F(i)} Z_v\right) \left( \bigotimes_{w\in P(i)} Z_w \right)  \left(\bigotimes_{u^\prime\in U(j)} X_{u^\prime}\right)\left(\mathbb{I}+ \bigotimes_{v^\prime\in F(j)} Z_{v^\prime}\right)  \left(\bigotimes_{w^\prime\in P(j)} Z_{w^\prime} \right)\\
  & = \; \frac{\left(-1\right)^{\theta_{ij}}}{4}  \left(\bigotimes_{u\in U(i)} X_u\right)\left(\bigotimes_{u^\prime\in U(j)} X_{u^\prime}\right) \left(\mathbb{I}- \bigotimes_{v\in F(i)} Z_v\right) \left( \bigotimes_{w\in P(i)} Z_w \right)\left(\mathbb{I}+  \bigotimes_{v^\prime\in F(j)} Z_{v^\prime}\right)  \left(\bigotimes_{w^\prime\in P(j)} Z_{w^\prime} \right)\\
    &= \; \frac{\left(-1\right)^{\theta_{ij}+\theta_{ji}}}{4} \left(\bigotimes_{u^\prime\in U(j)} X_{u^\prime}\right) \left(\mathbb{I}+ \bigotimes_{v^\prime\in F(j)} Z_{v^\prime}\right)  \left(\bigotimes_{w^\prime\in P(j)} Z_{w^\prime} \right)\left(\bigotimes_{u\in U(i)} X_u\right) \left(\mathbb{I}- \bigotimes_{v\in F(i)} Z_v\right) \left( \bigotimes_{w\in P(i)} Z_w \right)\\
    &\;\hat{=} \; - c^{\dagger}_j c^{\;}_i \; .
  \end{align}
  At last, we find by explicit construction :
  \begin{align}
  c^{\dagger}_jc^{\;}_j \; \hat{=} \; \frac{1}{2}\left( \mathbb{I}-\bigotimes_{v\in F(j)} Z_v \right)\, , \qquad  c^{\;}_j c^{\dagger}_j \; \hat{=} \; \frac{1}{2}\left( \mathbb{I}+\bigotimes_{v\in F(j)} Z_v \right)\, .
  \end{align}
  Thus,  we find $[c_i^{},c_j^{\dagger}]_{+}\;\hat{=}\;\delta_{ij}\mathbb{I}$, and our construction \eqref{eq:line} is in compliance with all relations in \eqref{eq:antiraw}.

\end{document}